\documentclass[aps,prd,onecolumn,showpacs,floatfix,preprintnumbers,amsfont,amsmath,amssymb,nofootinbib, superscriptaddress]{revtex4}
\usepackage{graphicx}% Include figure files
\usepackage{dcolumn}% Align table columns on decimal point
\usepackage{bm}% bold math
\usepackage{hyperref}% add hypertext capabilities
\usepackage{amsmath}
\usepackage{appendix}
\usepackage{amssymb}
\usepackage{color}
\usepackage{courier}
\usepackage{graphicx}
\usepackage{subcaption}
\graphicspath{{./}{figures/}}

\begin{document}

%\preprint{APS/123-QED}

\title{Convergence and Efficiency of Different Methods to Compute the Diffraction Integral for Gravitational Lensing of Gravitational Waves}

\author{Xiao Guo}%[0000-0001-5174-0760]
\email{guoxiao@nao.cas.cn}
\affiliation{CAS Key Laboratory for Computational Astrophysics, National Astronomical Observatories, Chinese Academy of Sciences, 20A Datun Road, Beijing 100101, China}
\affiliation{School of Astronomy and Space Science, University of Chinese Academy of Sciences, 19A Yuquan Road, Beijing 100049, China}
\author{Youjun Lu}%[0000-0002-1310-4664]
\email{luyj@nao.cas.cn}
\affiliation{CAS Key Laboratory for Computational Astrophysics, National Astronomical Observatories, Chinese Academy of Sciences, 20A Datun Road, Beijing 100101, China}
\affiliation{School of Astronomy and Space Science, University of Chinese Academy of Sciences, 19A Yuquan Road, Beijing 100049, China}

\date{\today}% It is always \today, today,

\begin{abstract}
Wave optics may need to be considered when studying the lensed waveforms of gravitational waves (GWs). However, the computation of the diffraction integral (amplification factor) in wave optics is challenging and time-consuming. It is vital to develop an accurate and efficient method to calculate the amplification factor for detecting lensed GW systems. In this paper, we investigate the convergence of the diffraction integral for gravitational lensing of GWs and analyze the accuracy and efficiency of a number of numerical methods that can be used to calculate this integral, including the integral mean method, asymptotic expansion method, {\bf Levin's} method, zero points integral method, etc. We further introduce a new method by combining the zero points integral and the asymptotic expansion methods to calculate the diffraction integral, which provides an efficient and accurate way to calculate the lensed waveform of GWs. %In addition, we also check the small-angle approximation adopted in the traditional diffraction integral for GW lensing and find it does not bring significant deviation from the true one.
\end{abstract}
%
%\keywords{Research Areas: Electromagnetic radiation astronomy - Gravitational lenses; Gravitational waves - Gravitational wave detection;
%Techniques: Computational techniques: Numerical approximation & analysis
%Gravitation, Cosmology & Astrophysics}
\maketitle

\section{Introduction}

The wave optics is frequently considered for the gravitational lensing of gravitational waves (GWs) because the GW wavelengths can be comparable or even larger than the Schwarzschild radius of lens objects (i.e., $\lambda\gtrsim R_{\rm S}$) in many cases, which provides accurate GW waveform estimates after the lensing \citep{1998PhRvL..80.1138N, 2003ApJ...595.1039T, 2014PhRvD..90f2003C, 2018PhRvD..98j4029D, 2019ApJ...875..139L}. \citet{1992grle.book.....S} and \citet{2003ApJ...595.1039T} derive the formulas for diffraction integral of the amplification factor by adopting the eikonal approximation, thin lens approximation, and the small angle approximation, in the wave optics regime. The wave optics is also needed when considering the intensity of lensed electromagnetic (EM) wave at the caustics because it is divergent in the geometrical optics limit \citep{1981Ap&SS..78..199B, 1986ApJ...307...30D, 1992grle.book.....S}, although geometrical optics is sufficient to describe the EM wave lensing in most realistic astrophysical cases as $\lambda\ll R_{\rm S}$.

Detection of the gravitational lensing of GW events needs accurate and efficient calculations of the lensed GW signal. The extraction of lensed GW signal is based on the matched filtering method, which is sensitive to the lensed GW waveform \citep{1998PhRvL..80.1138N, 2011gwpa.book.....C, 2014PhRvD..90f2003C, 2018PhRvD..98j4029D} and needs the construction of a template bank with a large number of lensed GW templates. Therefore, an accurate and efficient numerical integral method for calculating the diffraction integral would be vital for lensed GW detection. However, it is a difficult challenge to accurately compute the traditional diffraction integral derived for gravitational lensing of GWs (or EM waves) in the wave optics regime \citep{2003ApJ...595.1039T}. The reason is that it is a general (improper) integral defined on an infinite area and the integrand is a rapidly oscillating function, which is usually difficult to integrate by using ordinary numerical methods such as the Gaussian quadrature. It is time consuming to obtain accurate results by using ordinary integration methods since a little change of the settings for the upper limit of the integral leads to a large change in the integral value.
%
%\citet{longman_1956} provides a method to calculate the integral of oscillatory function on infinite interval. 
%Similar to it, 
In addition, the traditional diffraction integral is highly oscillatory even at infinity, and thus it is possibly not well-defined by means of the usual convergence definition (see calculation method of such integral introduced in \citep{longman_1956}). It is important to make it clear whether such an integral is slowly or weakly convergent. 

In this paper, we demonstrate the diffraction integral is Ces\`{a}ro summable, which means it is convergent in the mean. And we show that the Ces\`{a}ro sum of traditional diffraction integral is basically consistent with the general diffraction integral without the small-angle approximation. This suggests that the small-angle approximation is a good approximation and the traditional diffraction integral formulas can compute the lensed waveform with high accuracy. Then we overview a number of numerical methods to calculate the diffraction integral that have been introduced in the literature, including the time delay contour integration method \citep{1995ApJ...442...67U}, asymptotic expansion method \citep{Press1992NumericalRI, Takahashi_Thesis}, {\bf Levin's} method \citep{Levin1982}, Filon-type method \citep{filon_1930,Xiang2007}, etc. (see for a summary in \citep{Iserles2006, Iserles2006b, 2008mgm..conf..807M}). 
We further introduce several new methods like the integral mean, zero points integral, and zero points-asymptotic expansion methods to calculate this integral. We explicitly demonstrate the validity of these new methods, and compare them with other methods in the literature on the aspects of convergence, accuracy, efficiency, etc.

This paper is organized as follows. In Section~\ref{sec:WOB}, we derive the general formulas of diffraction theory and discuss the accuracy of the traditional diffraction integral. In Section~\ref{sec:conver}, we investigate the convergence of the diffraction integrals, prove it is convergent in the mean. Then we introduce several numerical methods to compute diffraction integrals and compare them in Section \ref{sec:numerical}. Conclusions and discussions are summarized in Section \ref{sec:concl}. 

Throughout the paper, we adopt the geometrical unit system $G=c=1$.

\section{Wave Optics}
\label{sec:WOB}
{\bf
Based on \cite{2003ApJ...595.1039T}, under the eikonal approximation, GW tensor can be described as a scalar wave 
$$h_{\mu \nu}=\phi(t) e_{\mu \nu},$$
where $e_{\mu\nu}$ is the polarization tensor, $\phi(t)$ represents the GW waveform in time domain. We use $\tilde{\phi}$ to represent the waveform in the frequency domain as the Fourier transform of $\phi$. 

\begin{figure}
\centering
\includegraphics[width=0.5\textwidth]{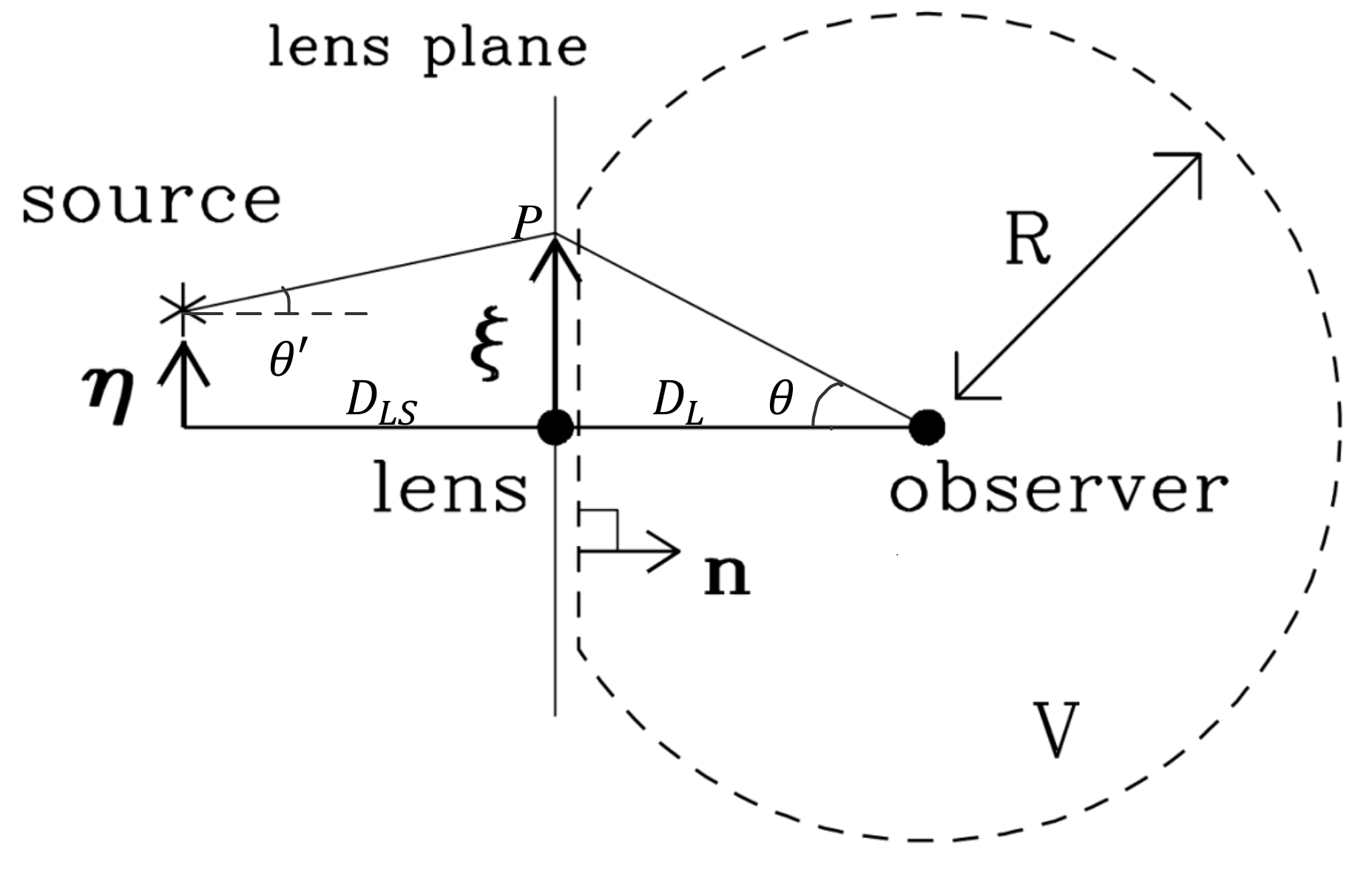}
\caption{
Cartoon diagram to illustrate the geometrical configuration of the observer-lens-GW source system (modified from \cite{Takahashi_Thesis}). Here $D_{\rm L}$ and $D_{\rm S}$ represent the distances from the observer to the lens and source, respectively, and $D_{\rm LS}$ represents the distance between the lens and source; $\boldsymbol{\eta}$ and $\boldsymbol{\xi}$ are the position vectors of source on the source plane and GW on the lens plane, respectively; $\boldsymbol{n}$ is the normal vector of the lens plane. The volume $V$ is enclosed by a spherical surface with radius $R$ and the lens plane but does not include lens.
}
\label{fig:illustr}
\end{figure}

For the source-lens-observer system, we use $D_{\rm S}$ ($D_{\rm L}$) to represent the distance between source (lens) and observer, and $D_{\rm LS}$ the distance between source and lens. As shown in Figure~\ref{fig:illustr},
the GW emitted from a source at a position  $\boldsymbol{\eta}$ in the source plane reaches point $P$ ($\boldsymbol{\xi}$) on the lens plane and is finally received by the distant observer. 
The distance between point $P$ and observer is $r$, $\theta$ and $\theta'$ represent the angle between the normal vector of lens plane and GW propagation direction at the position of observer and source, respectively, $\omega$ represents the circular frequency of GW. It can be proven that the observed lensed waveform in the frequency domain $\tilde{\phi}_{\rm obs}^{L}$ propagating in a curved spacetime can be expressed as a surface integral at lens plane 
}
\begin{equation}
\tilde{\phi}_{\rm obs}^{L}=\frac{1}{4 \pi} \iint_{S} d^{2} \xi\left[\tilde{\phi} \left(-i \omega \cos \theta \frac{e^{i \omega r}}{r}\right)
-\frac{e^{i \omega r}}{r} i \omega \cos \theta^{\prime} \tilde{\phi}\right].
\label{eq:phi_incl}
\end{equation}
{\bf where scalar wave at lens plane can be expressed as $\tilde{\phi}=Ae^{iS_{\rm p}}$, and
$A$ and $S_{\rm p}$ represent the amplitude and phase of GW.
More details about the derivation are described in Appendix~\ref{sec:deriv}.
}
Due to $r=D_{\rm L}/\cos\theta$, we have
\begin{eqnarray}
\tilde{\phi}_{\rm obs}^{\rm L} 
& = & \frac{\omega A}{4 \pi i D_{\rm L}} \iint_{S} d^{2} \xi e^{i \omega t_{\rm d}}(\cos \theta+\cos\theta')\cos\theta,
\label{eq:phi_cos}
\end{eqnarray}
where $t_{\rm d}$ is the time delay, $\cos\theta= D_{\rm L}/\sqrt{D_{\rm L}^2+\xi^2}$ and $\cos\theta'=D_{\rm LS}/\sqrt{D_{\rm LS}^2+(\boldsymbol{\xi}-\boldsymbol{\eta})^2}$.

The amplification factor of lens can be defined as
\begin{equation}
F(\omega, \boldsymbol{\eta})=\frac{\tilde{\phi}_{\rm obs}^{\rm L}(\omega, \boldsymbol{\eta})}{\tilde{\phi}_{\rm obs}(\omega, \boldsymbol{\eta})},
\end{equation}
where $\tilde{\phi}_{\rm obs}=\frac{D_{\rm LS}A}{D_{\rm S}}e^{iS_{\rm P0}}$, thus we have
\begin{equation}
F(\omega, \boldsymbol{\eta})=\frac{D_{\rm S}}{D_{\rm L} D_{\rm LS}} \frac{\omega}{4 \pi i} \iint_S d^{2} \xi e^{i \omega t_{\rm d}(\boldsymbol{\xi}, \boldsymbol{\eta})}(\cos \theta+\cos\theta')\cos\theta.
\label{eq:fweta}
\end{equation}
where the phase $S_{\rm P0}$ is absorbed into $t_{\rm d}$.
By defining dimensionless quantities, i.e., $\boldsymbol{x}=\frac{\boldsymbol{\xi}}{\xi_{0}}, \boldsymbol{y}=\frac{D_{\rm L}}{\xi_{0} D_{S}} \boldsymbol{\eta}$, $w=\frac{D_{\rm S}}{D_{\rm LS} D_{\rm L}} \xi_{0}^{2}\left(1+z_{\rm L}\right)\omega$,  $T(\mathbf{x}, \mathbf{y})=\frac{D_{\rm L} D_{\rm LS}}{D_{\rm S}} \xi_{0}^{-2} t_{\rm d}(\boldsymbol{\xi}, \boldsymbol{\eta})$, with $\xi_0$ as a normalized constant of length, Equation~\eqref{eq:fweta} can be reduced to 
\begin{equation}
F(w, \boldsymbol{y})=\frac{w}{4 \pi i} \iint_S d^{2} x e^{i w T(\boldsymbol{x}, \boldsymbol{y})} (\cos\theta+\cos\theta') \cos\theta.
\label{eq:F_wycos}
\end{equation}
Here $\cos\theta=\frac{\frac{D_{\rm L}}{\xi_0}}{\sqrt{\left(\frac{D_{\rm L}}{\xi_0}\right)^2+x^2}}$, $\cos\theta'=\frac{\frac{D_{\rm LS}}{\xi_0}}{\sqrt{\left(\frac{D_{\rm LS}}{\xi_0}\right)^2+\left(\boldsymbol{x}-\frac{D_{\rm S}}{D_{\rm L}}\boldsymbol{y}\right)^2}}$ and
\begin{equation}
T(\boldsymbol{x},\boldsymbol{y})=\frac{1}{2}|\boldsymbol{x}-\boldsymbol{y}|^{2}-\psi(\boldsymbol{x})+\phi_{m}(\boldsymbol{y}),
\label{eq:Time}
\end{equation}
with $\psi(\boldsymbol{x})$ representing the lens potential, $\phi_m(\boldsymbol{y})$ representing the arrival time in the unlensed case.

\subsection{Axial Symmetric Case}

In polar coordinate system $(x,\theta_x)$, where $\theta_x$ is the angle between $\boldsymbol{x}$ and $\boldsymbol{y}$,
Equation~\eqref{eq:F_wycos} can be written as
\begin{equation}
F(w, \boldsymbol{y})=\frac{w}{4 \pi i} \int_0^{\infty} xd x\int_0^{2\pi} d\theta_x \exp\left[i w \left(\frac{x^2}{2}+\frac{y^2}{2}-xy\cos\theta_x-\psi(\boldsymbol{x})+\phi_{\rm m}(\boldsymbol{y}\right)\right] (\cos\theta+\cos\theta') \cos\theta,
\label{eq:F_wycospolar}
\end{equation}
where
\begin{equation}
\cos\theta'=\frac{\frac{D_{\rm LS}}{\xi_0}}{\sqrt{\left(\frac{D_{\rm LS}}{\xi_0}\right)^2+x^2+\frac{D^2_{\rm S}}{D^2_{\rm L}}y^2-2\frac{D_{\rm S}}{D_{\rm L}}xy\cos\theta_x}}.
\end{equation}
If the lens mass distribution is axial symmetric, $\psi(\boldsymbol{x})=\psi(x)$ is independent of $\theta_x$.

\subsection{Traditional Form from Small Angle Approximation}

If we adopt the small-angle approximation, i.e., $\cos\theta\simeq1$, $\cos\theta'\simeq1$, $r = D_{\rm L}/\cos \theta \simeq D_{\rm L}$, which should be proper for most astrophysical lensing systems, then Equation~\eqref{eq:phi_cos} can be reduced to 
\begin{equation}
\tilde{\phi}_{\rm obs}^{\rm L}(\omega, \boldsymbol{\eta})=\frac{\omega A}{2 \pi i D_{\rm L}} \iint d^{2} \xi \exp \left[i \omega t_{\rm d}(\boldsymbol{\xi}, \boldsymbol{\eta})\right].
\end{equation}
Similarly, amplification factor can be expressed as
\begin{equation}
F(w, \boldsymbol{y})=\frac{w}{2 \pi i} \iint d^{2} x \exp [i w T(\boldsymbol{x}, \boldsymbol{y})].
\label{eq:F_wy}
\end{equation}
This formula is widely used in the calculation of amplification factor in the wave optics regime (e.g., see \citep{2003ApJ...595.1039T,2014PhRvD..90f2003C, 2018PhRvD..98j4029D, 2019ApJ...875..139L,2019A&A...627A.130D,2020PhRvD.101l3512D}). In the axial symmetric case,
\begin{equation}
\begin{aligned}
F(w, \boldsymbol{y})
=&\frac{w}{2 \pi i} e^{iw(y^2/2+\phi_m(\boldsymbol{y}))}\int_0^{\infty} xdx e^{iw(x^2/2-\psi(x))}\int_0^{2\pi} d\theta_x e^{-i w xy\cos\theta_x} \\
=&\frac{w}{i} e^{iw(y^2/2+\phi_m(\boldsymbol{y}))}\int_0^{\infty} xdx e^{iw(x^2/2-\psi(x))}J_0(wxy),
\end{aligned}
\label{eq:axial}
\end{equation}
where the Bessel function $$J_0(z)=\frac{1}{\pi}\int_0^{\pi}e^{iz\cos\theta}d\theta. $$

\subsection{Accuracy of the Small Angle Approximation}
\label{sec:acc}

Since $\frac{D_{\rm L}}{\xi_0}$ and $\frac{D_{\rm LS}}{\xi_0}$ are usually large numbers, we can expand cosine functions into series to estimate the errors of Equation \eqref{eq:F_wy}. When $x$, $y$ are relatively small, $\frac{D_{\rm L}}{\xi_0}$, $\frac{D_{\rm LS}}{\xi_0}\gg x, y$, we can expand $\cos\theta$ and $\cos\theta'$ in the form of series and we have
\begin{equation}
\frac{(\cos\theta+\cos\theta')\cos\theta}{2}\approx1-\frac{3}{4}\left(\frac{x}{\frac{D_{\rm L}}{\xi_0}}\right)^2-\frac{1}{4}\left(\frac{\boldsymbol{x}-\frac{D_{\rm S}}{D_{\rm L}}\boldsymbol{y}}{\frac{D_{\rm LS}}{\xi_0}}\right)^2+...
\end{equation}
If we regard Equation~\eqref{eq:F_wycos} as the standard results, this expansion can also give an estimate to the relative error of the traditional Equation~\eqref{eq:F_wy}, which is roughly
\begin{equation}
\epsilon=\mathcal{O}\left(\left(\frac{x}{\frac{D_{\rm L}}{\xi_0}}\right)^2\right)+\mathcal{O}\left(\left(\frac{|\boldsymbol{x}-\frac{D_{\rm S}}{D_{\rm L}}\boldsymbol{y}|}{\frac{D_{\rm LS}}{\xi_0}}\right)^2\right).
\end{equation}

Usually, normalization length is taken as Einstein radius $\xi_0=r_{\rm E}=\sqrt{\frac{4 G M_{\rm L}}{c^{2}} \frac{D_{\rm LS}D_{\rm L}}{D_{\rm S}}}=\sqrt{2R_{\rm S}\frac{D_{\rm LS}D_{\rm L}}{D_{\rm S}}}$, where $R_{\rm S}$ the Schwarzschild radius of the lens object. For typical lens systems, $D_{\rm LS}\sim D_{\rm L} \sim D_{\rm S}$ and $\frac{D_{\rm L}}{\xi_0}\sim\frac{D_{\rm LS}}{\xi_0}\sim\frac{D_{\rm L}}{\sqrt{R_{\rm S}D_{\rm L}}}=\sqrt{\frac{D_{\rm L}}{R_{\rm S}}}$. Therefore, the relative error of the integrand in Equation~\eqref{eq:F_wy} is $\sim (x^2+|\boldsymbol{x}-\boldsymbol{y}|^2)\frac{R_{\rm S}}{D_{\rm L}}$. 

For general lens systems, distance $D_{\rm L}\sim D_{\rm LS}\sim1.7$\,Gpc ($z\sim1$, assuming the concordance flat $\Lambda$CDM cosmology), lens mass $M_L\lesssim 10^{12}M_\odot$  \citep{2018MNRAS.476.2220L, 2018PhRvD..98j4029D}, thus these two ratios $\frac{D_{\rm L}}{\xi_0}\simeq\frac{D_{\rm LS}}{\xi_0}\simeq\sqrt{\frac{D_{\rm L}}{R_S}}\gtrsim 10^5$. The error is only the order of $\mathcal{O}\left[(\frac{x}{10^5})^2\right] \lesssim 10^{-10}$, which is negligible. There might be some extreme rare cases, for example, the lens system is a nearby galaxy/dark matter halo with distance $D_{\rm L} \sim 1$\,Mpc and the GW source is at $z\sim 1$, or the lens system is a galaxy/dark matter halo close to the GW source ($D_{\rm LS} \sim 1$\,Mpc) and the GW soure is at $z\sim 1$, in which one of the two ratios ($\frac{D_{\rm L}}{\xi_0}, \frac{D_{\rm LS}}{\xi_0}$) is possibly as small as $\sim10^3$. In such cases, the difference between Equation~\eqref{eq:F_wy} and Equation~\eqref{eq:F_wycos} can be estimated. As an example, we adopt the singular isothermal sphere (SIS) lens model (see  \citep{2003ApJ...595.1039T}) to estimate this difference. We use $F_1$ to represent the traditional diffraction integral (Equation~\eqref{eq:F_wy}), $F_2$ to represent the general diffraction integral (Equation~\eqref{eq:F_wycos}), thus $|F_1|-|F_2|$ is the module difference between two integrals and $\arg(F_1)-\arg(F_2)$ is the phase difference. When $y=10$, $w\sim0.1$, we obtain $|F_1|-|F_2|\sim 10^{-4}$, $\arg(F_1)-\arg(F_2)\sim 10^{-5}$. Therefore, the waveform error induced by the small angle approximation would be  $\lesssim \delta h/h\sim 10^{-4}$ for most cases. Such an error is negligible since it cannot be detected ($\langle \delta h|\delta h\rangle<1$ \citep{Lindblom:2008cm}) when the SNR $\lesssim 10^4$, which is satisfied for most GW detection. 
Therefore, we conclude here that the traditional diffraction integral derived from the small angle approximation is sufficiently accurate to derive the amplification factor for lensed GW signals.

\section{Convergency of Integral}
\label{sec:conver}

In this section, we first discuss the convergence of the diffraction integrals (Eqs.~\eqref{eq:F_wy} and \eqref{eq:F_wycos}) and  demonstrate the traditional diffraction integral Equation~\eqref{eq:F_wy} is $(C,1)$ summable. 
The meanings about $(C,1)$ summable and Ces\`{a}ro Summability are described in Appendix~\ref{sec:Ces_sum}.

\subsection{Convergence Analysis}

The traditional diffraction integral Equation~\eqref{eq:F_wy} appears not absolutely convergent when $x\rightarrow\infty$. The term $\exp[iwT(x,y)]$ is a highly oscillating function, and the ``amplitude" of the oscillation does not decay with increasing upper limit of the integral $x_{\rm u}$. In the axial symmetric case, when $x\rightarrow\infty$, ${\rm Amp}(J_0(x))\propto x^{-0.5}$ which is always oscillating between $\pm x^{-0.5}$, where $\rm Amp(\cdots)$ represents the amplitude of the oscillation of $\cdots$, and thus $\lim_{x\rightarrow\infty} xJ_0(x)=\infty$. The other term $e^{iw(x^2/2-\psi(x))}$ is usually rapidly oscillating with increasing $x$ to large values. However, this seeming divergence or singularity can be removed by variable substitution $z=\frac{x^2}{2}$, which will be discussed in Section~\ref{sec:IntMean} in detail.

Taking the NFW model 
\cite{NFW1997}
as an example, we set 
$\psi(x)=\frac{\kappa}{2}(\ln^2\frac{x}{2}-{\rm arctanh}^2(\sqrt{1-x^2}))$\footnote{It should be noticed that when $x>1$, $-{\rm arctanh}^2\sqrt{1-x^2}$ is replace by ${\rm arctan}^2\sqrt{x^2-1}$ in complex field.} 
with $\kappa=1$ and $y=0$, 
where $\kappa$ is the dimensionless mass surface density parameter \citep{2004A&A...423..787T}, 
then we can obtain
\begin{equation}
|F|=w\left|\int_0^{x_{\rm u}}xdxe^{iw(x^2/2-\psi(x))}\right|,
\label{eq:Fabs_xu}
\end{equation}
where $x_u$ is the upper limit of the integral.

\begin{figure}
\centering
\includegraphics[width=0.5\textwidth]{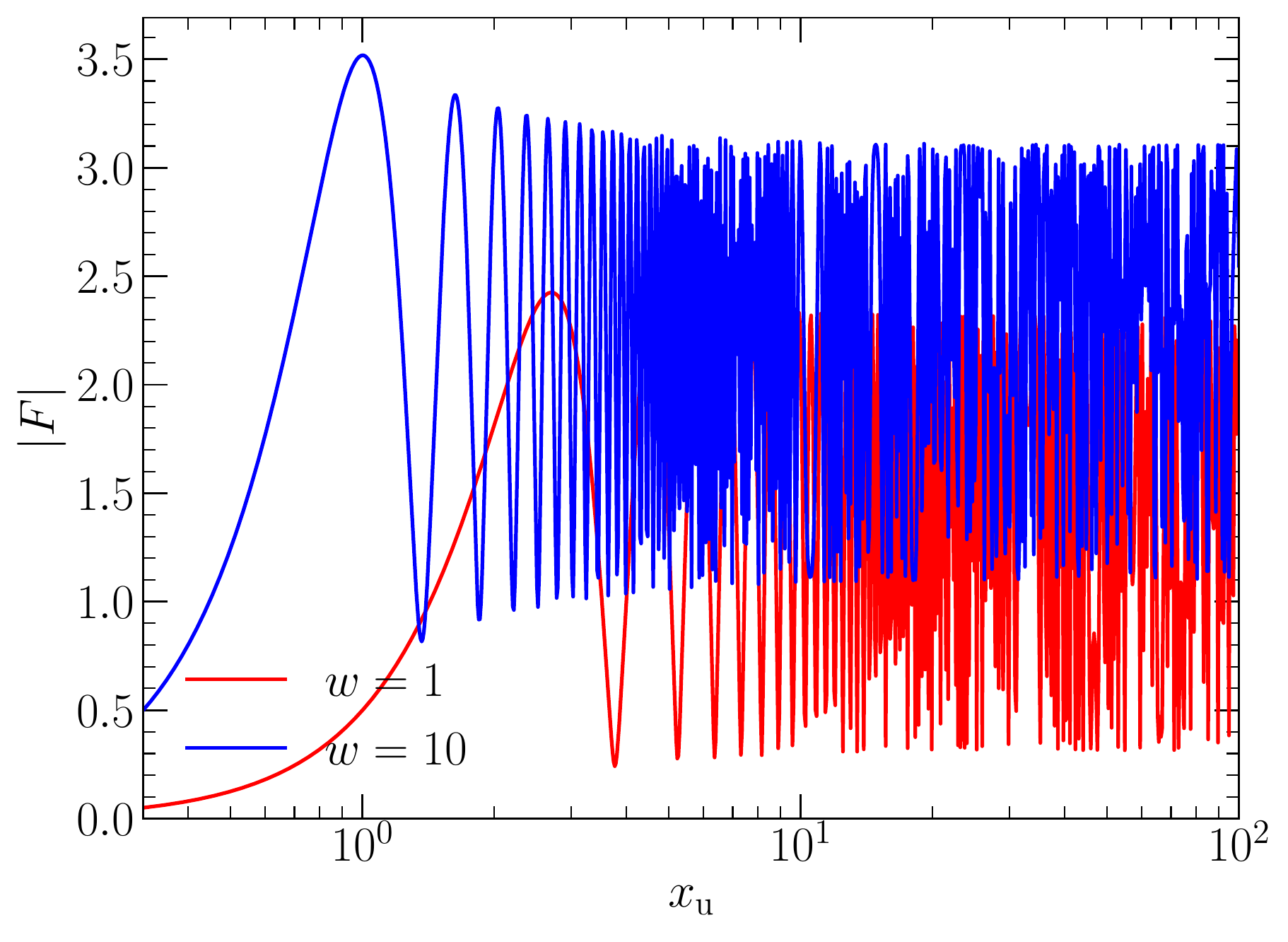}%wave_optics test.ipynb
\caption{The $|F|$ as the function of $x_{\rm u}$. The integral value is rapidly oscillatory with increasing upper limit $x_{\rm u}$. Even it is convergent at infinity, it must be slowly convergent.
}
\label{fig:oscillatory}
\end{figure}

Figure~\ref{fig:oscillatory} shows the integral $|F|$ (Eq.~\eqref{eq:Fabs_xu}) as a function of the upper limit $x_{\rm u}$ for two different $w$, which is a highly oscillatory function. Such an oscillating behaviour of the integral at $x_{\rm u}\rightarrow \infty$ suggests that it could be slowly or weakly convergent. Although this integral is not convergent under the usual convergence definition, we can still generalize the definition of convergence for it as long as it is $(C,1)$ summable, i.e., it is convergent to a value $L$ in the mean, and thus $L$ can be defined as the integral value of Equation~\eqref{eq:F_wy}. As shown in Section~\ref{sec:trad_sum}, the traditional diffraction integral Equation~\eqref{eq:F_wy} is usually $(C,1)$ summable and we obtain a convergent integral value $L$ when we use any special integration method to numerically calculate such a highly oscillatory improper integral.

Similar to the traditional diffractional integral form (Eq.~\eqref{eq:F_wy}), the general diffraction integral (Eq.~\eqref{eq:F_wycos}) is also a rapidly oscillatory when $x$ is small. However, when $x\rightarrow\infty$, $\cos\theta \propto 1/x$, $\cos\theta'\propto 1/x$, and thus $\exp [i w T(\boldsymbol{x}, \boldsymbol{y})] (\cos\theta+\cos\theta') \cos\theta\propto \exp[iwT(x,y)] /x^{2}$, where $\exp [i w T(\boldsymbol{x}, \boldsymbol{y})]$ is bounded above. In this case, a finite constant $C$ can always be found so that $\exp [i w T(\boldsymbol{x}, \boldsymbol{y})](\cos\theta+\cos\theta')\cos\theta <Cx^{-2}$. Thus the integral $\int d^{2} x \exp [i w T(\boldsymbol{x}, \boldsymbol{y})](\cos\theta+\cos\theta')\cos\theta< \pi C$ is absolutely convergent even at infinity. Thus this improper integral can be well-defined.

\subsection{The Ces\`{a}ro Summability of Traditional Diffraction Integral}
\label{sec:trad_sum}

Equation~\eqref{eq:F_wy} can be expressed as
\begin{equation}
F(w, \boldsymbol{y})=\frac{w}{2 \pi i} \iint d^{2} x \{\cos [ w T(\boldsymbol{x}, \boldsymbol{y})]+i\sin [ w T(\boldsymbol{x}, \boldsymbol{y})]\}
\label{eq:F_R_I}
\end{equation}
The geometrical meaning of this 2D integral is a complex volume $V_{\rm tot}=V_{\rm tot}^{\rm c}+iV_{\rm tot}^{\rm s}$, where $V_{\rm tot}^{\rm c}$ ($V_{\rm tot}^{\rm s}$) represents the volume of the real (imaginary) part of the integral. $V_{\rm tot}^\alpha$ (where $\alpha=$c, s) can be express as the positive volume $V_+^\alpha$ (the part with integrand greater than zero) minus the negative volume $|V_-^\alpha|$ (the part with integrand less than zero), similar to that in \citet{longman_1956}. 

According to that the zero curves $l^{\rm c}_{n}$ (or $l^{\rm s}_{n}$) satisfy $wT=(n+\frac{1}{2})\pi$ (or $wT=n\pi$) for $\cos wT$ (or $\sin wT$), we can divide the total volume $V_{\rm tot}^\alpha$ into infinite parts $V^{\rm c}_{n}$ (or $V^{\rm s}_{n}$) where $V^{\rm c}_{n}$ (or $V^{\rm s}_{n}$) satisfy $(-\frac{1}{2}+n)\pi<wT<(\frac{1}{2}+n)\pi$ (or $n\pi<wT<(n+1)\pi$), with $n=0,1,2,3,\cdots$. Since $wT\geq 0$, $V_{0}^{\rm c}$ actually represents the volume on the zone where $0<wT<\frac{1}{2}\pi$. For $\cos wT$ (or $\sin wT$), when $(-\frac{1}{2}+2k)\pi<wT<(\frac{1}{2}+2k)\pi$ (or $2k\pi<wT<(2k+1)\pi$), the volume $V^{\rm c}_{2k}$ ($V^{\rm s}_{2k}$) is positive; when $(\frac{1}{2}+2k)\pi<wT<(\frac{3}{2}+2k)\pi$ (or $(2k+1)\pi<wT<(2k+2)\pi$), the volume $V^{\rm c}_{2k+1}$ (or $V^{\rm s}_{2k+1}$) is negative, here $k=0,1,2,3,\cdots$. Thus 
$$
V_{\rm tot}^{\alpha}=\sum_{n=0}^{\infty}V_{n}^{\alpha}=\sum_{n=0}^{\infty}(-1)^{n}|V_{n}^{\alpha}|.
$$
Hence we can then transform the improper integral into two alternate series $\sum_{n=0}^{\infty}V_{n}^{\alpha}$. The improper integral is $(C,1)$ summable to $V_{\rm tot}$, i.e., series $\sum_{n=0}^{\infty}V_{n}^{\alpha}$ is $(C,1)$ summable to $V_{\rm tot}^\alpha$ for both $\alpha=$ c, s.

According to Equation~\eqref{eq:Time}, $T(x) \propto x^2$ when $x \rightarrow \infty$ since $\psi(x)$ is usually much smaller than $x^2$ (e.g., $\psi(x) \simeq \ln x$ for the point mass model, or $x$ for the SIS model; \citealt{Takahashi_Thesis})\footnote{For other lens potentials $\psi(x)$, see \cite{2001astro.ph..2341K}. If $\psi(x)$ grows faster than the geometrical time delay $\frac{x^2}{2}$ with increasing $x$, this integral may be divergent. However, such a case is extremely rare. 
If $\psi(x)\rightarrow x^{\beta}$ when $x\rightarrow\infty$, where $\beta$ represents the index of the power law. As long as the total mass of lens is finite, we should have {\bf $\beta<0$. For example, the total mass of SIS model with $\beta=1$ has been divergent due to its mass-radius relation $M(r)\propto r$.} }. 
We set $T(x)\simeq C_Tx^2$ ($C_T>0$) when $x\rightarrow\infty$, $V_n^{\rm c}$ (or $V_n^{\rm s}$) is defined on the zone $S_n^{\rm c}$ (or $S_n^{\rm s}$) where $(n-\frac{1}{2})\pi<wC_Tx^2<(n+\frac{1}{2})\pi$ (or $n\pi<wC_Tx^2<(n+1)\pi$), i.e.  $(n-\frac{1}{2})\pi<\Phi<(n+\frac{1}{2})\pi$ (or $n\pi<\Phi<(n+1)\pi$), here we set $\Phi=wC_{T}x^2$. Thus we have
$$
|V_n^{\rm c}|\approx\left|\iint_{S_n^{\rm c}}d^2x\cos(wC_Tx^2)\right|=\left|\frac{\pi}{wC_T}\int_{(n-\frac{1}{2})\pi}^{(n+\frac{1}{2})\pi} d\Phi \cos(\Phi)\right|=\frac{2\pi}{wC_T},
$$
when $n$ is substantially large. Similarly, we also have
$$
|V_n^{\rm s}|\approx\frac{2\pi}{wC_T},
$$
when $n$ is a large integer. Therefore, $|V_{n}^{\alpha}|\rightarrow\frac{2\pi}{wC_T}$ for both $\alpha=$\,c and s when $n>m$ and $m$ is a large integer. The series 
$$
V_{\rm tot}^{\alpha}=\sum_{n=0}^{\infty}V_{n}^{\alpha}=\sum_{n=0}^mV_n^{\alpha}+\sum_{n=m+1}^{\infty}(-1)^n|V_{n}|
=\sum_{n=0}^mV_n^\alpha+\sum_{n=m+1}^{\infty}(-1)^n\frac{2\pi}{wC_T},
$$
where $\alpha=$ c or s, which means the partial sum $P_k^\alpha=\sum_{n=0}^kV_n^{\alpha}$ is  oscillating with $k$ with an amplitude of $\frac{2\pi}{wC_T}$ even at $k\rightarrow\infty$. When $k\leq m$, $P_k^\alpha=\sum_{n=0}^kV_n^{\alpha}$; when $k>m$, $P_k^\alpha=P_m^\alpha+\frac{\pi}{wC_T}[(-1)^{k}-(-1)^m]$. Therefore $V_{\rm tot}^{\alpha}$ is always oscillating with an amplitude of  $\frac{2\pi}{wC_{T}}$, and possibly not convergent. However, the above demonstration is not strict since many approximations were adopted and we can not draw the conclusion that the diffraction integral must be divergent at infinity. 
Above proof assumes that the integral are axisymmetric thus $\int_0^{2\pi}d\theta_x=2\pi$, actually, $\int_0^{2\pi}d\theta_xe^{-iwxy\cos\theta_x}=2\pi J_0(wxy)\leq\int_0^{2\pi}d\theta_x=2\pi$ become smaller and smaller with increasing $x$ and basically always less than $2\pi$. Thus the "amplitude" of oscillation is actually not constant, but become weaker and weaker. Due to ${\rm Amp}(J_0(x))\propto x^{-0.5}$, we have $|V_{n}^{\alpha}|\approx\frac{2\pi}{wC_T}J_0(wxy) \propto\frac{2\pi}{wC_T\sqrt{wxy}}$ when $x\rightarrow\infty$. As long as the integral is convergent to a value $L$, it is easy to prove that it must be $(C,1)$ summable to the same value $L$.\cite{NIST:DLMF}
Nevertheless, as an approximation, this at least tells us that the diffraction integral must be slowly convergent although it is not divergent.

Even if $V_{\rm tot}^{\alpha}$ is divergent, we can still  redefine 
\begin{eqnarray*}
V_{\rm tot}^{\alpha}&\equiv&\lim_{n\rightarrow\infty}\sum_{k=0}^{n}\frac{P_{k}^{\alpha}}{n+1}
=\lim_{n\rightarrow\infty}\left(\sum_{k=0}^{m}\frac{P_{k}^{\alpha}}{n+1}+\sum_{k=m+1}^{n}\frac{P_m^\alpha+\frac{\pi}{wC_{T}}[(-1)^{k}-(-1)^m]}{n+1}\right)\\
&=&P_{m}^{\alpha}+\frac{\pi}{wC_{T}}(-1)^m=\sum_{n=0}^mV_n^\alpha+\frac{\pi}{wC_{T}}(-1)^m.
\end{eqnarray*}
This limit exists and can be well-defined. According to Equation~\eqref{eq:Time}, we usually have the constant $C_{T}=\frac{1}{2}$ when there is no shear. Shear could change the coefficient $C_{T}$ \citep{1995ApJ...442...67U}. Thus the integral Equation~\eqref{eq:F_R_I} is $(C,1)$ summable to
\begin{equation}
V_{\rm tot}=\sum_{n=0}^mV_n^{\rm c} +\frac{2\pi}{w}(-1)^m+i\left[\sum_{n=0}^mV_n^{\rm s} +\frac{2\pi}{w}(-1)^m\right],
\label{eq:V_tot}
\end{equation}
where $m$ is a large integer. If one wants to study the deviation of $T\propto x^2$, one can regard $C_T$ as the polynomial expansion of $1/x$: $C_T(x)=C_{T0}(1+c_0/x+\cdots)$.  

In conclusion, the integral Equation~\eqref{eq:F_R_I} is $(C,1)$ summable to Equation~\eqref{eq:V_tot} as long as $T(x)\propto x^2$ when $x\rightarrow\infty$. In practical numerical calculations, $V_{\rm tot}^{\alpha} \approx \sum_{k=0}^{m} P_{k}^{\alpha}/ (m+1)$ can always be served as a good numerical estimation for $V_{\rm tot}^\alpha$ even if $T(x)\propto x^2$ is not satisfied when $x\rightarrow\infty$, or the constant $C_{T}$ is not known.

\section{Numerical Integration Methods}
\label{sec:numerical}

The diffraction integral is difficult to directly compute as discussed above (see Figure~\ref{fig:oscillatory}). A numerical integration method is needed to calculate the amplification factor for most lens models, except that for the point mass lens model the amplification factor can be expressed analytically (see \citet{2003ApJ...595.1039T} and \citet{Takahashi_Thesis}). Below we overview and introduce a number of numerical integration methods to calculate this oscillatory integral. In the Ulmer \& Goodman\cite{1995ApJ...442...67U}'s method, the location of geometrical optics images are required to know at first and the line integration on constant time delay may be difficult to compute. We do not expand the discussion on this method as it may be not so convenient in generating lensed GW template bank with all kinds of parameters and lens models.

\subsection{Integral Mean Method}
\label{sec:IntMean}
According to the definition of $(C,1)$ summability, a rapidly oscillatory improper integral $I=\int_{0}^{\infty}g(z)dz$ can be calculated by using its integral mean, i.e.,
\begin{equation}
\int_{0}^{\infty}g(z)dz=\lim_{b\rightarrow\infty}I_{C}(b),
\end{equation}
where 
$$I_C(b)=\int_{0}^{b}\left(1-\frac{z}{b} \right)g(z)dz$$ 
(see Appendix \ref{sec:Ces_sum}), and the factor $1-\frac{z}{b}$ weakens the amplitude of the oscillation of this integral at large $z$.

In reality, we can use $I_{C}(b)$ to approximate $I$, where $b$ is such a large number that the fluctuation of $I_{C}(b)$ is within the error precision. We can also discretize the integral mean into $\bar{I}(z_j)=\sum_{i=1}^{j}I(z_i)$, where $I(z_i)=\int_0^{z_i}g(z)dz$ and $z_j$ is a large number. However, this algorithm may be not so efficient like $I_{C}(b)$.

In the axial symmetric case, we make the transform $z= x^2/2$, according to Equation~\eqref{eq:axial}, and thus
\begin{eqnarray*}
F(w, \boldsymbol{y})
=\frac{w}{i} e^{iw(y^2/2+\phi_m(\boldsymbol{y}))}\int_0^{\infty} dz e^{iw(z-\psi(\sqrt{2z}))}J_0(wy\sqrt{2z}).
\end{eqnarray*}
For the denotation of $I_C(z)$, we have 
$$
g(z)=e^{iw(z-\psi(\sqrt{2z}))}J_0(wy\sqrt{2z}).
$$
For this integral
$$
I(\infty)=\int_0^{\infty}e^{iw(z-\psi(\sqrt{2z}))}J_0(wy\sqrt{2z})dz,
$$
where the oscillation ``amplitude" of the integrand $g(z)$ goes to zero at infinity. This integral represents the area $A$ closed in $g(z)$ with $z$ axis. (When $g(z)<0$, the area is negative.) The zero points of integrand $g(z)$ divide the total area $A$ into many small area $A_i$ ($A_i$ can be negative). As long as $\sum_{i=1}^{\infty}A_i$ is convergent, this integral is convergent.

In the non-axial symmetric case, $\psi(\boldsymbol{x})=\psi(x,\theta_x)$,
\begin{equation}
F(w, \boldsymbol{y})
=\frac{w}{2 \pi i} e^{iw(y^2/2+\phi_m(\boldsymbol{y}))}\int_0^{\infty} dz e^{iwz}\int_0^{2\pi} d\theta_x e^{-i w(\psi(\sqrt{2z},\theta_x)+\sqrt{2z}y\cos\theta_x)}, 
\end{equation}
thus
$$
g(z)=e^{iwz}\int_0^{2\pi} d\theta_x e^{-i w (\psi(\sqrt{2z},\theta_x) +\sqrt{2z}y\cos\theta_x)} 
$$
for $I_C(z)$.

For illustration, the blue lines in Figures~\ref{fig:diff_int} and \ref{fig:diff_int_p} show  two examples of the amplitude and phase of the amplification factor obtained from the integral mean method for the NFW lens model and the point mass lens model, respectively. As seen from these two Figures, the integral mean method works well in the aspect of convergence, though the convergence is not as faster as those with the asymptotic expansion method.

\subsection{Asymptotic Expansion Method}

The diffraction integral can be calculated by using the asymptotic expansion method \citep{Press1992NumericalRI,Press1997Numerical,Takahashi_Thesis,CHOIch2}. An improper integral with oscillatory integrand, similar to the diffraction integral, may be expressed as
\begin{eqnarray}
\int_{0}^{+\infty} d z e^{i \omega z} f(z)&=&\int_{0}^{b} d z e^{i \omega z} f(z)+\int_{b}^{\infty} d z e^{i \omega z} f(z)\\
&=&\int_{0}^{b} d z e^{i \omega z} f(z)+\frac{e^{i \omega z}}{i\omega} f(z)\left|_b^{\infty}\right.-\int_b^{\infty} \frac{e^{i \omega z}}{iw} \frac{\partial{f(z)}}{\partial z} dz\\
&=&\int_{0}^{b} d z e^{i \omega z} f(z)+\left.e^{i \omega b} \sum_{n=1}^{\infty} \frac{(-1)^{n}}{(i \omega)^{n}} \frac{\partial^{n-1} f}{\partial z^{n-1}}\right|_{z=b}
\end{eqnarray}

Due to that $e^{i\omega z}$ is always oscillating, this method assumes that integrand $f(z)$ and its any $n$-order derivative $\partial^n f(z)/\partial z^n$ go to zero at infinity \citep{Press1992NumericalRI,CHOIch2}. This condition may be difficult to be satisfied in the usual definition for diffraction integral on infinite area, however, under the definition of $(C,1)$ summability,
$$
\lim_{z\rightarrow\infty}\frac{e^{iwz}}{iw}f(z)\equiv\lim_{b\rightarrow\infty}\frac{1}{b}\int_{0}^{b}dz\frac{e^{iwz}}{iw}f(z).
$$
Since $e^{iwz}$ is oscillating between positive and negative values, the positive parts and negative parts will nearly counteract. As long as the integral $\int_0^{b}\frac{e^{iwz}}{iw}f(z)$ is always finite even when $b\rightarrow\infty$, the above limit will go to zero. For its derivatives, we can also draw similar conclusions.

According to Equation~\eqref{eq:axial}, in the axial symmetric case, $f(z)$ in the asymptotic expansion method is
$$
f(z)=e^{-iw\psi(\sqrt{2z})}J_0(wy\sqrt{2z}).
$$
Since $J_0(x)\rightarrow0$ when $x\rightarrow\infty$, $\partial^n f(z)/\partial z^n\rightarrow0$ at infinity is easy to be satisfied.
In the non-axial symmetric case, $f(z)$ is a little complex and it is
$$
f(z)=\int_0^{2\pi} d\theta_x e^{-i w (\psi(\sqrt{2z},\theta_x) +\sqrt{2z}y\cos\theta_x)}. 
$$

\begin{figure}
\centering
\includegraphics[width=\textwidth]{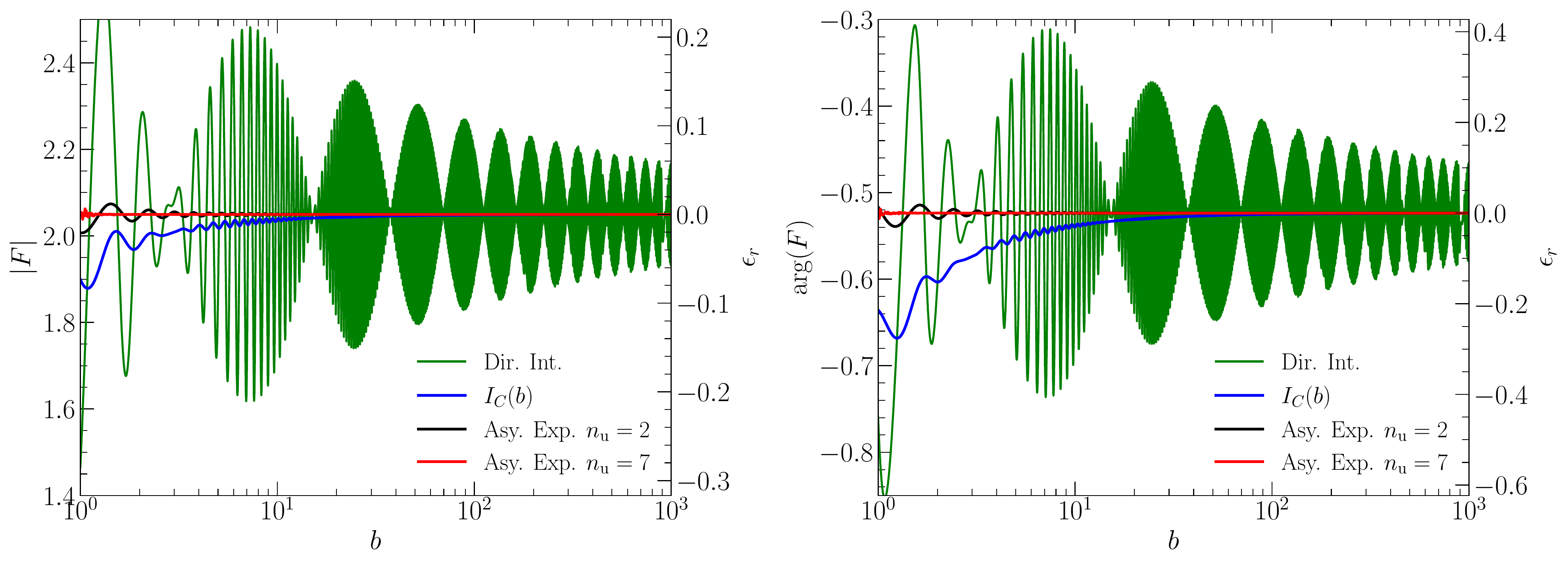}%Asymptotic_expansion test.ipynb
\caption{The amplification factor as a function of the upper limit $b$ of the integration range for the NFW lens model, with $y=0.1$, $w=10$, and $\kappa=1$. Left and right panels show the module $|F|$ and phase $\arg(F)$ of the amplification factor as the function of $b$, respectively. The right vertical axis ($\epsilon_{\rm r}$) indicates the relative error of $|F|$ (left) or $\arg(F)$ (right), i.e. $\epsilon_{\rm r}=\frac{|F|-|F_t|}{|F_t|}$ (left) or $\epsilon_{\rm r}=\frac{\arg(F)-\arg(F_t)}{|\arg(F_t)|}$ (right), where $|F_t|\approx2.0495$ or $\arg(F_t)\approx-0.5237$ represents the referenced true value obtained by the computation. For simplicity, here we set $\phi_m(y)\equiv0$. The green line represents the direct integration of $I(b)$ by using the Gauss quadrature. The blue line represents the results $I_{C}(b)$ obtained by using the integral mean method. The red and black lines represent the results obtained by using the asymptotic expansion method with $n_{\rm u} =2$ and $7$, respectively. 
}
\label{fig:diff_int}
\end{figure}

The asymptotic expansion method contains the priori assumption that these integrals are all convergent at infinity, so that additional terms can be obtained to suppress the oscillation of the integral as shown in Figure~\ref{fig:diff_int}. %This assumption may be incorrect in some cases, however it can give a convergent value consistent with the integral mean, if the diffraction integral is convergent in the mean.

One may have to set an upper limit for $n$ ($n_{\rm u}$ rather than $n=\infty$) in practical calculations using the asymptotic expansion method and assume that the higher order derivatives with $n>n_{\rm u}$ can be ignored. 
\begin{equation}
I_{{\rm A.E.}}(b)=\int_{0}^{b} d z e^{i w z} f(z)+\left.e^{i w b} \sum_{n=1}^{n_{\rm u}} \frac{(-1)^{n}}{(i w)^{n}} \frac{\partial^{n-1} f}{\partial z^{n-1}}\right|_{z=b}
\label{eq:AsyExp}
\end{equation}
The error of $I_{\rm A.E.}(b)$ as the approximation of $I(\infty)$ is $\mathcal{O}\left(e^{iwb}(i/w)^{n_{\rm u}+1} f^{({n_{\rm u}})}(b)\right)\rightarrow0$ when $b\rightarrow\infty$, especially when $w$ is large. In principle, choosing a larger $n_{\rm u}$ may lead to faster convergence of this integral to some extent. However, we find that the function is almost most rapidly convergent when $n_{\rm u}=7$  in our practical calculation. For a larger $n_{\rm u}$, this method may lead to a divergent result, which may be caused by the numerical errors in the calculations of higher order derivatives with $n>7$. The black and red lines in Figure~\ref{fig:diff_int} show the numerical results of the magnitude (right panel) and phase (right panel) of the amplification factor obtained by using the asymptotic expansion method with $n_{\rm u}=2$ and $7$, respectively, for the NFW lens model, while the black and red lines in Figure~\ref{fig:diff_int_p} correspondingly show the results for the point mass lens model. As seen from both Figures, the asymptotic expansion method offers a rapid convergence of the diffraction integral with increasing $b$ in practical calculations.   

\subsection{{\bf Levin's} method}
\label{sec:Levin}

{\bf Levin's} method can be used to compute such oscillatory integral $$I(b) = \int_{a}^{b} f(x)e^{iq(x)}dx,$$ where $f(x)$ is a non-oscillatory function \citep{Levin1982}.
{\bf Levin's} method for 1-dimensional integral has three steps:
\begin{enumerate}
\item Choosing linearly independent base functions $u_k(x)$, $k=1,2,\cdots,n$, like polynomials, or Chebyshev polynomials and so on, whose properties are simliar to $f(x)$.
\item Solving $\alpha_{k}$ from collocation equations $$\sum_{k=1}^{n} \alpha_{k} u_{k}^{\prime}\left(x_{j}\right)+i q^{\prime}\left(x_{j}\right) \sum_{k=1}^{n} \alpha_{k} u_{k}\left(x_{j}\right)=f\left(x_{j}\right),$$ where $x_j=a+(j-1)(b-a)/(n-1)$, and $j=1,2,\cdots,n$.
\item Computing $I_n(b)$ as the approximation to $I(b)$ $$I_{n}(b)=\sum_{k=1}^{n} \alpha_{k} u_{k}(b) e^{i q(b)}-\sum_{k=1}^{n} \alpha_{k} u_{k}(a) e^{i q(a)}$$.
\end{enumerate}

\begin{figure}
\centering
\includegraphics[width=\textwidth]{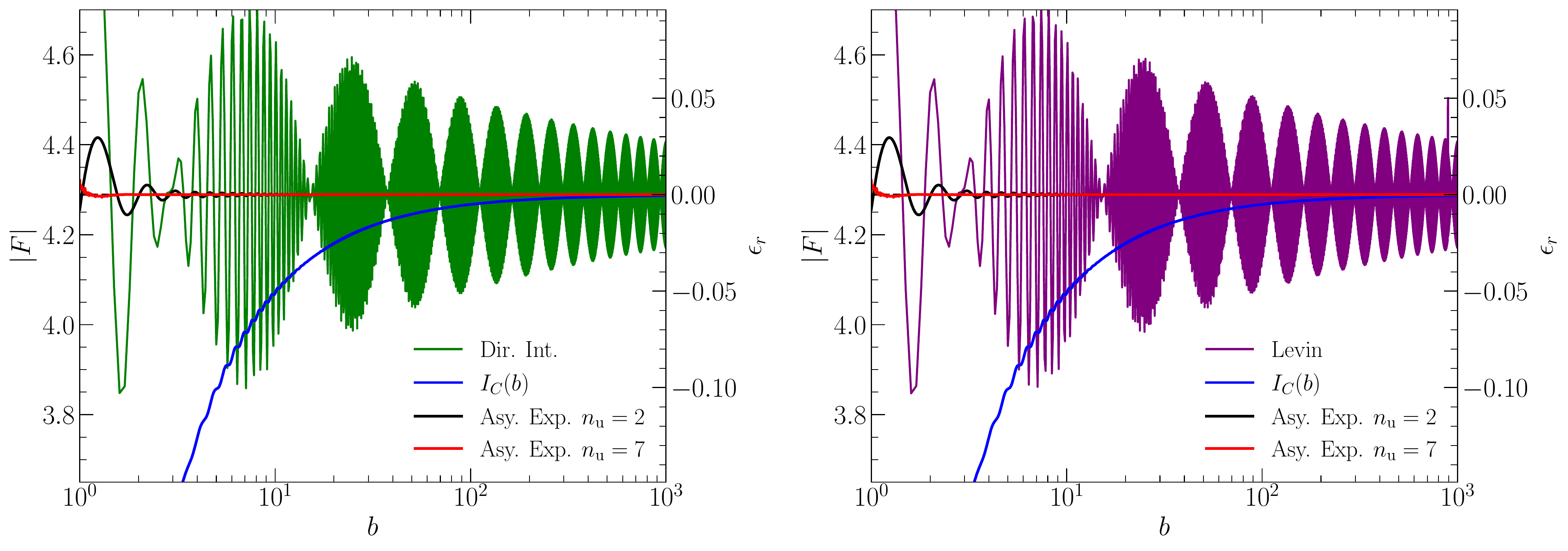}%wave_optics test.ipynb
\caption{The module of the amplification factor as a function of the upper limit $b$ of the integration range for the point mass lens model, with $y=0.1$, $w=10$. The right vertical axis ($\epsilon_{\rm r}$) indicates the relative error of $|F|$, i.e. $\epsilon_{\rm r}=\frac{|F|-|F_p|}{|F_p|}$, where $|F_p|$ represents the true value obtained by the computation of analytical expression.  The green line in the left panel represents the direct integration by using the Gaussian quadrature. The blue line in each panel represents the integral mean $I_C(b)$. The black and red solid lines in each panel are obtained by using the asymptotic expansion method with $n_{\rm u}=2$ and $7$, respectively. The purple line in the right panel is obtained by using the {\bf Levin's} method. They all converge to the value of the analytical expression for the point mass lens model (black dashed horizontal line; see \citep{2003ApJ...595.1039T}). The black dashed lines are covered by the red line. Only the module $|F|$ is shown here. For the phase, it is similar to that for the module, as shown in Fig.~\ref{fig:diff_int}. 
}
\label{fig:diff_int_p}
\end{figure}

For {\bf Levin's} method, it just transforms an arbitrary non-oscillatory function $f(x)$ into the linear combinations of many independent base functions $u_{k}(x)$, $k=1,2,\cdots,n$ and their derivatives. Then it can transform the integral into the difference between the summation of linear combinations of base functions times $e^{iq(x)}$ at the upper and lower limits ($b$ and $a$) of the integral. When $b$ is finite, this method is easy to realize. The purple line in the right panel of Figure~\ref{fig:diff_int_p} shows an example obtained by using the {\bf Levin's} method, where the base function $u_k(x)$ are Chebyshev polynomials. The integral value is basically consistent with the direct integration, which is slowly convergent. However, when $b\rightarrow\infty$, we cannot make sure that the base functions on infinite interval $\lim_{b\rightarrow\infty}\sum_{k=0}^{n}\alpha_{k}u_k(b)e^{iq(b)}$ are convergent in the usual definition. For example, if the base functions are polynomials, it is obvious divergent in the usual integration definition because  $u_{k}(x)\rightarrow\infty$ when $x\rightarrow\infty$. However, according to $(C,1)$ summability, it is sufficient to have
\begin{equation}
\lim_{b\rightarrow\infty}\overline{\sum_{k=0}^{n}\alpha_{k}u_k(b)e^{iq(b)}}=\lim_{b\rightarrow\infty}\frac{\int_{0}^{b}\sum_{k=0}^{n}\alpha_{k}u_k(x)e^{iq(x)}dx}{b}=0,
\label{eq:Levin_base}
\end{equation}
as long as the mean below is convergent. Although we cannot make sure the above limit must be convergent, which may depend on the choices of the base functions and $q(x)$, this convergent condition is weaker than the usual definition. Since $e^{iq(x)}$ is rapidly oscillating between positive and negative values when $x$ is large, most parts of the integral vanish just like an alternate series as long as $u_{k}(x)$ does not grow up  fast.

For the diffraction integral to be computed, $a=0$. Combining with the integral mean, we can generalize the Levin method to the integral on infinite intervals, i.e., 
\begin{equation}
I(\infty)\approx 0-\sum_{k=1}^{n} \alpha_{k} u_{k}(0) e^{i q(0)}.
\end{equation}
We can choose a large $b$ as an approximation to it. 

In addition, one may make variable substitution to transform the infinite interval into finite interval. Although the integral after transformation may be defined on a finite interval, it is still rapidly oscillating near the singularity. Therefore, the variable substitution cannot solve the problem of oscillatory integral. 

\subsection{Zero Points Integral Method}

The diffraction integral over infinite interval can be also computed by using the method in \citet{longman_1956}, which transforms the integral into a series summation. Then some mathematical methods may be used to accelerate the convergence of series to obtain the integral value such as the Euler's transformation of series \citep[See ][]{NIST:DLMF}. The speed of convergence partly depends on the choice of the method to accelerate the convergence of series. However, for this zero points integral method, it may be difficult to find the exact positions of all zero lines of the integrand, especially for $2$-dimensional diffraction integrals. For some specific integrands of the diffraction integral like the integrands consist of $J_{0}(x)$, this method can be improved accordingly and thus can provide an efficient way to compute the integral.

For the diffraction integral in axial symmetric case, our aim is to compute this integral $I(\infty)$, where 
$$
I(b)\equiv\int_0^{b} dz e^{iw(z-\psi(\sqrt{2z}))}J_0(wy\sqrt{2z})
$$
When $y\neq0$, we need to find out the zero points of $J_0(wy\sqrt{2z})$. We set the $k$-th zero points of $J_0(x)$ is $x=j_k$, $k=1,2,3,\cdots$. Thus the zero points of $J_0(wy\sqrt{2z})$ are $z_k=\frac{j_k^2}{2w^2y^2}$. Within the neighborhood of $z_k$, the ``amplitude" of the oscillation of this integral is the smallest, thus the error of evaluation of the integral value would be the smallest as long as the error of the position of zero point $j_k$ is accurate. We just need to compute these integrals $I(z_k)$, which can be computed by the Guassian quadrature or {\bf Levin's} method, thus we have
\begin{equation}
I(\infty)=\lim_{k\rightarrow\infty}I(z_k)
\label{eq:Ics}
\end{equation}
as long as $I(\infty)$ does exist. Even when $I(\infty)$ does not exist or it is slowly convergent, we can redefine the $(C,1)$ sum of this integral (See Appendix~\ref{sec:Ces_sum}) as
\begin{equation}
I_{CS}(\infty)=\lim_{m\rightarrow\infty}\sum_{k=1}^{m}\frac{ I(z_k)}{m},
\end{equation}
where the adoption of the $(C,1)$ sum $I_{CS}(z_m)=\sum_{k=1}^{m}\frac{ I(z_k)}{m}$ can accelerate the convergence of $I(z_n)$. This would be an efficient method to compute $I(\infty)$.
Nevertheless, when $k$ is large, for example, $k=20$, the $20-$th zero point of $J_0(z)$ is $z_{20}\approx1925$ in our later examples, this method is possible to required to calculate the integral $I(z_{20})$, where $z_{20}$ is quite large. When $b$ is a large number, the calculation of $I(b)$ may be time consuming and have a large error.

\subsection{Zero Points Asymptotic Expansion Method}
\label{sec:zpae}

In order to avoid computing integral $I(z_{k})$ at large $z_{k}$, we introduce a new method by combining the zero points integral method and the asymptotic expansion method together, which we denote it as the zero points asymptotic expansion method. It only needs to compute the integral $I(z_k)$ when $k$ is a small integer. As for this new method, we have
\begin{equation}
I(\infty)=I(z_k)+ \left.e^{i \omega z_k} \sum_{n=1}^{\infty} \frac{(-1)^{n}}{(i \omega)^{n}} \frac{\partial^{n-1} f}{\partial z^{n-1}}\right|_{z=z_k}.
\end{equation}

We can use $I_{\rm A.E.}(z_k)$ as the approximation of $I(\infty)$. If we combine $I_{CS}(z_k)$ with asymptotic expansion method, we have
\begin{equation}
I_{\rm CS,A.E.}(\infty)=\lim_{m\rightarrow\infty}\sum_{k=1}^{m}\frac{ I_{\rm A.E.}(z_k)}{m},
\end{equation}
here we can discard the first one or two zero points if they have relative large errors.
This method is not only as accurate as other methods but also much more efficient than other methods.

\begin{figure}
\centering
\includegraphics[width=\textwidth]{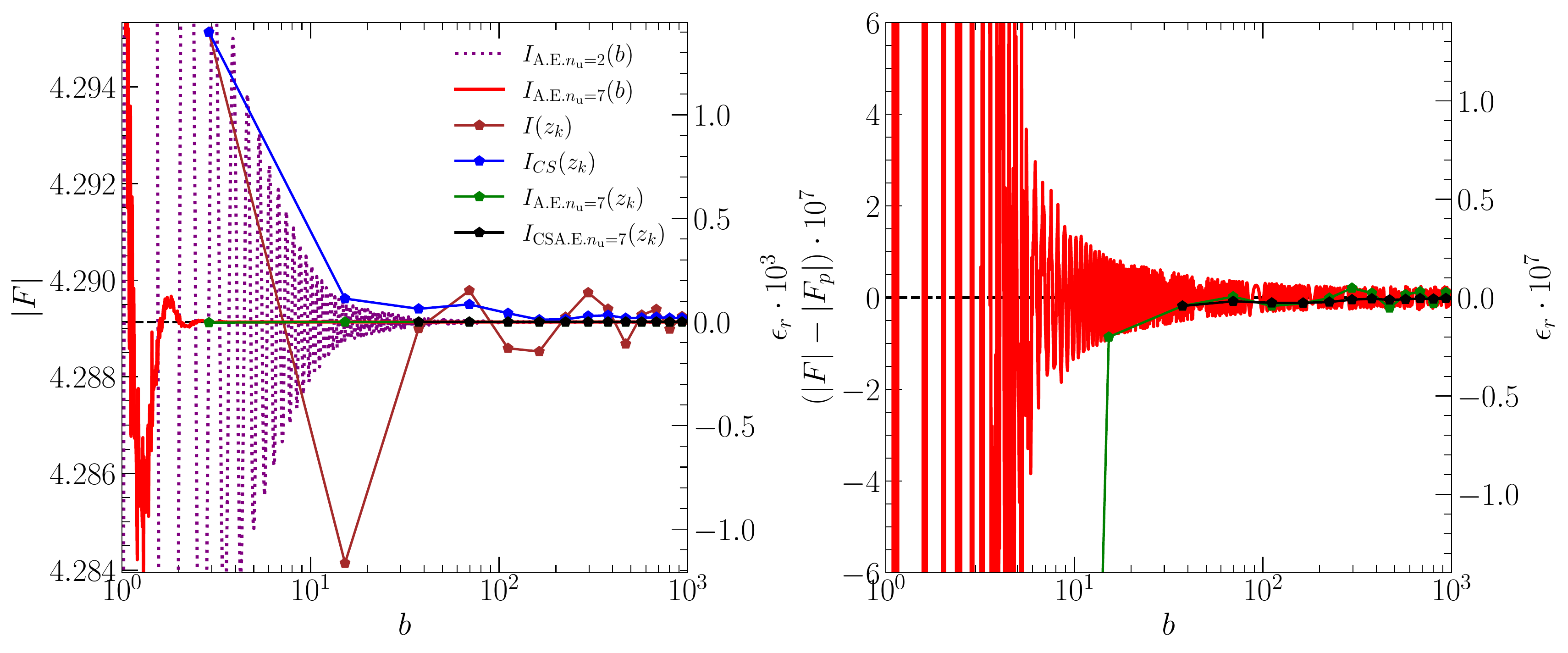}%wave_optics test.ipynb
\caption{Left panel: amplitude of the amplification factor $|F|$ as a function of the upper limit $b$ for the integration range, obtained by using several different methods for the point mass lens model with $y=0.1$, $w=10$. The purple dot line and the red solid line represent the results obtained by using the asymptotic expansion method with $n_{\rm u}=2$ and $7$, respectively. The brown points, blue points, green points, and black points represent the results obtained by using the zero points integral method ($I(z_k)$), the average of the zero points integrals $I_{CS}(z_k)$, the zero points asymptotic expansion method $I_{\rm A.E.} (z_k)$ with $n_{\rm u} =7$ and its mean $I_{\rm CS, A.E.} (z_k)$, respectively. The horizontal black dash line represents the analytical result (see \citep{2003ApJ...595.1039T}). Right panel: accuracy of the amplitude of the amplification factor $|F|-|F_{\rm p}|$ calculated by using different methods. $|F_{\rm p}|$ represents the true value obtained by the computation of analytical expression. 
The legend is the same as for the left panel. Here we discard the first two points which have large errors for the black line.
}
\label{fig:point_new}
\end{figure}

\begin{figure}
\centering
\begin{subfigure}[b]{0.48\textwidth}
\includegraphics[width=\textwidth]{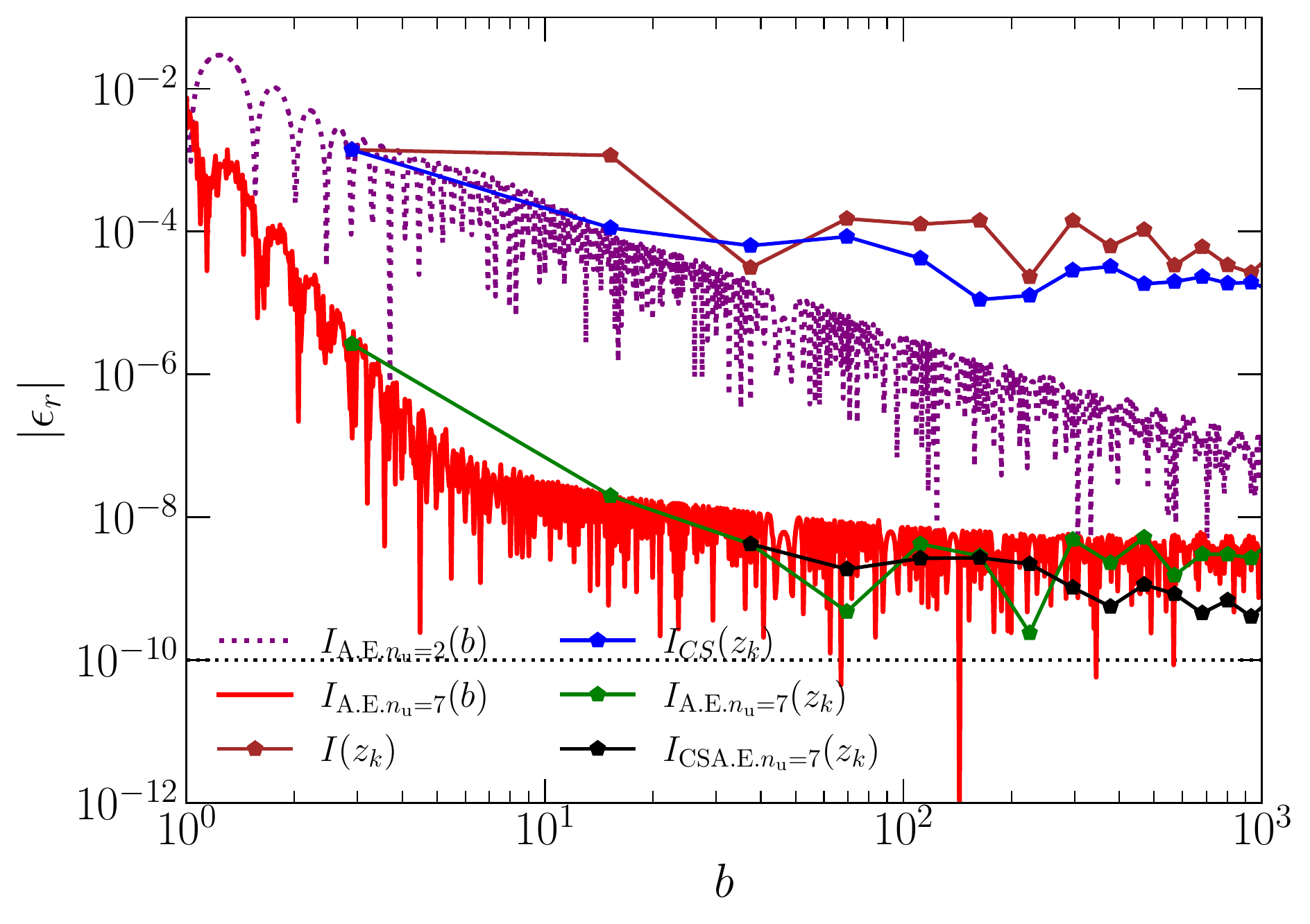}%wave_optics test.ipynb
\end{subfigure}
\hfill
\begin{subfigure}[b]{0.48\textwidth}
\includegraphics[width=\textwidth]{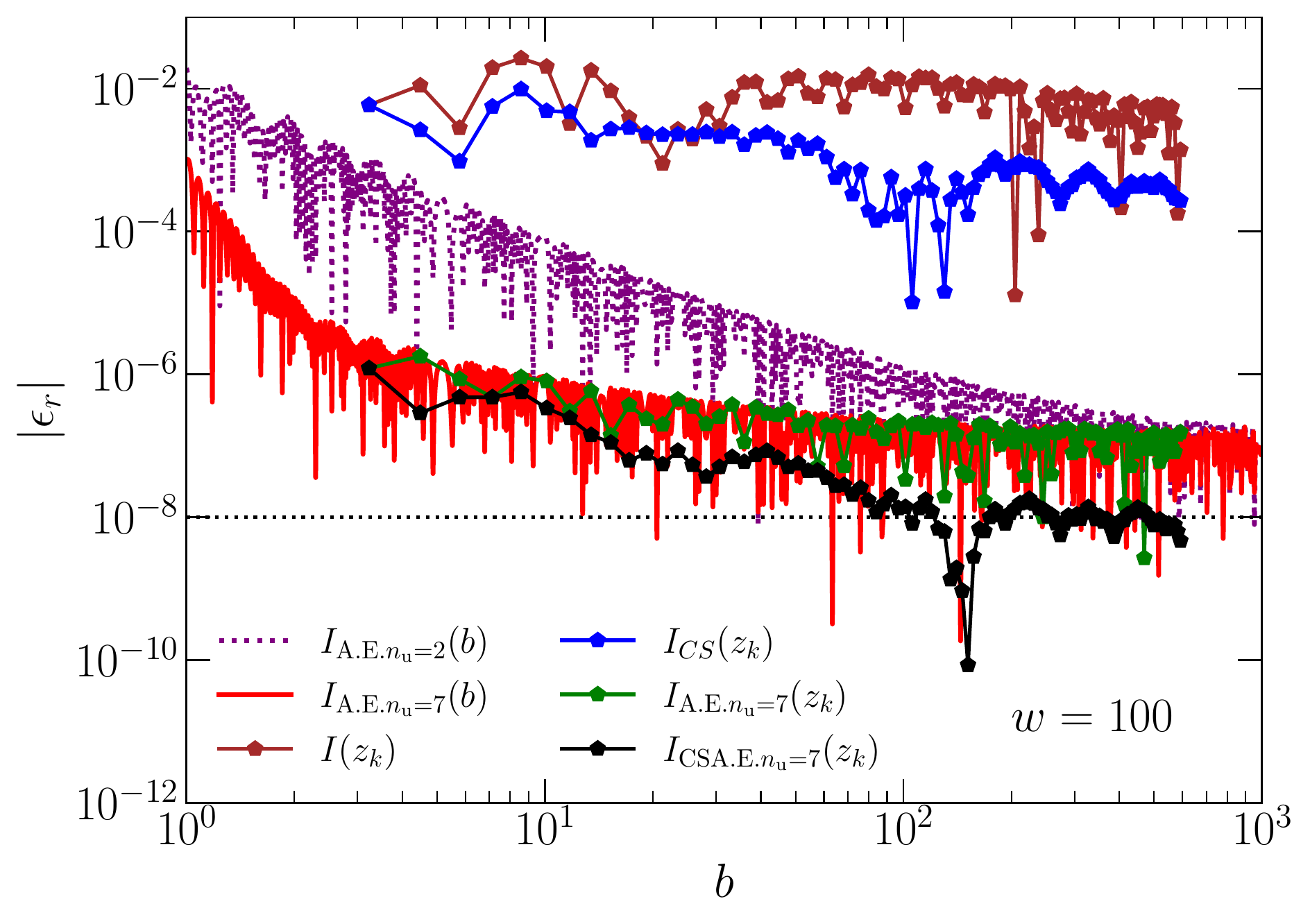}
\end{subfigure}
\caption{
The absolute value of the relative error $|\epsilon_r|$ of $|F|$ as a function of the upper limit $b$ for the integration range in log-log diagram, obtained by using several different methods for the point mass lens model with $y=0.1$, $w=10$ ($w=100$) in the left (right) panel. The horizontal black dot line represents the precision set for the numerical integration in our calculation. 
Curves and symbols with different colors/types represent the methods the same as those in Figure~\ref{fig:point_new}.
%The legend is the same as for  Figure~\ref{fig:point_new}.
%Here we discard the first two points which have large errors in the left panel. 
We choose the $20$-th to $100$-th zero points in the right panel. The relative errors of $I_{{\rm CS A.E.}n_{\rm u}=7}(z_k)$ become much smaller after averaging tens of points.
}
\label{fig:point_log}
\end{figure}

\begin{figure}
\centering
\includegraphics[width=\textwidth]{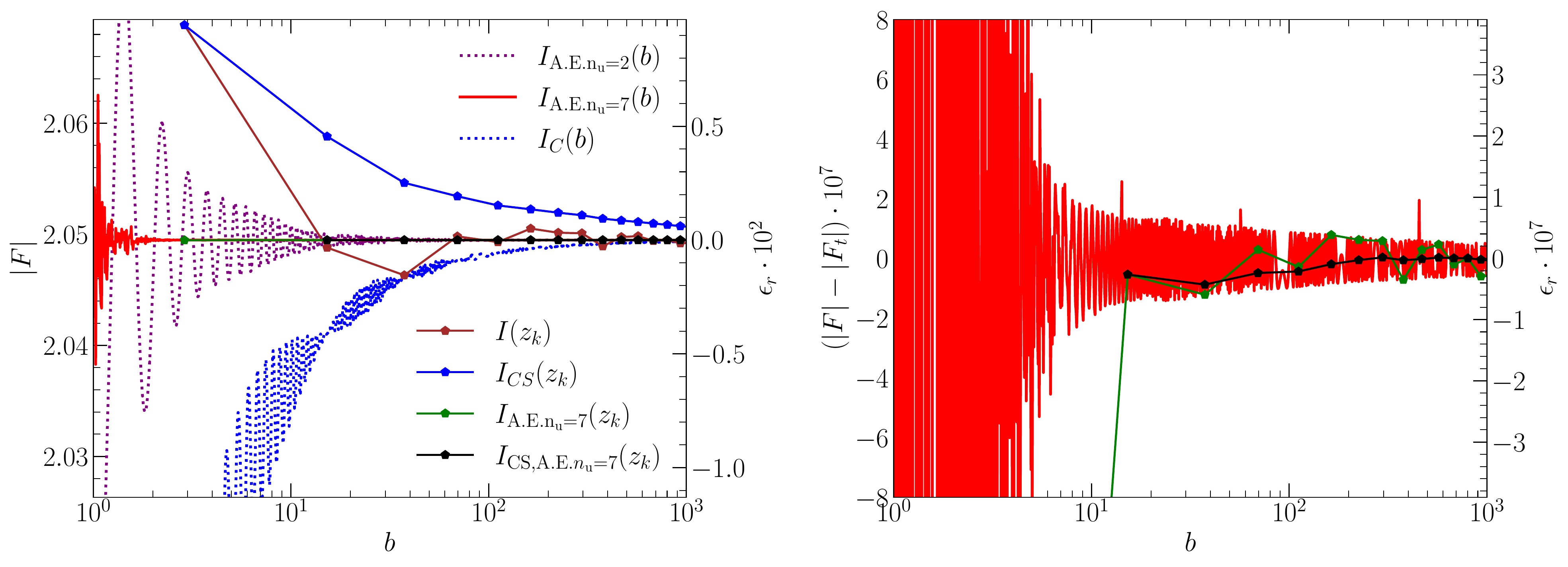}%Asymptotic_expansion test.ipynb
\caption{Legend similar to Figure~\ref{fig:point_new}, but for the NFW lens model, which does not have an analytical expression for the integral value and a reference value is taken as $|F_t|=2.049479253200136$. We also show the integral mean $I_C(b)$ in this figure as its deviation is relatively small in this case.
}
\label{fig:NFW_new}
\end{figure}

Taking the point mass lens model and the NFW lens model as two examples, here we calculate the amplification factor to investigate in the aspects of convergence, accuracy, and efficiency for these new methods based on zero points integral $I(z_k)$, $I_{CS}(z_k)$, and the combination of $I(z_k)$ ($I_{CS}(z_k)$) with the asymptotic expansion method, i.e., the zero points asymptotic expansion methods $I_{\rm A.E.}(z_k)$ ($I_{\rm CS,A.E.}(z_k)$). 
Figures~\ref{fig:point_new} and \ref{fig:NFW_new} show the results obtained by using the zero points asymptotic expansion method and its comparison with those from other methods for the point mass lens model and the NFW lens model, respectively. Figure~\ref{fig:point_log} shows the relative errors in log-log plot for point mass model, with which  the relative errors of different methods can be seen more clearly. Although the accuracy of the zero points integral method is not as high as the asymptotic expansion method at the same $z_k$, it is much more efficient as it only needs to calculate the integral values at a limited number of (zero) points to reach the same accuracy. The main reason is that it discards unnecessary integral values which bring violent oscillation, and thus enables a much more efficient way to obtain high computational accuracy, which will be further discussed in Section~\ref{sec:comp}. 

For $2$-dimensional integral $I_2(\infty)$, where
$$
I_2(b)\equiv\int_0^{b} dz e^{iwz}\int_0^{2\pi} d\theta_x e^{-i w(\psi(\sqrt{2z},\theta_x)+\sqrt{2z}y\cos\theta_x)},
$$
we can also find out the zero points of $I_{\theta_x}(z)\equiv\int_0^{2\pi} d\theta_x e^{-i w(\psi(\sqrt{2z},\theta_x)+\sqrt{2z}y\cos\theta_x)}$ as long as they exist. To figure out their zero points, we only need to calculate out the zero points of real function $|I_{\theta_x}(z)|$. This integral is not difficult to compute as long as $w$ and $z$ are not very large. The time to find out these zero points is short, which is much less than the time to compute 2 dimensional integral. We have tested it for some cases. It really works.

\subsection{Summary for the Comparisons of Different Methods}
\label{sec:comp}

In this Section, we summarize our main results by comparing different numerical methods in calculating the diffraction integral introduced above and comment on its convergence and efficiency.

We show the integral results by using the direct integration, the asymptotic expansion method, and the integral mean method to compute the traditional diffraction integral with parameters $(w,y)=10,0.1$ in Figure~\ref{fig:diff_int} for the point mass lens model and Figure~\ref{fig:diff_int_p} for the NFW lens model, respectively. The green lines show the integral values obtained from the direct integration of $I(b)$ with the Gauss quadrature method, which are rapidly oscillating and slowly (or just $(C,1)$) convergent with increasing $b$. The blue lines show the integral values of $I_{C}(b)$ obtained by using the integral mean method and the Gauss quadrature method, which are not highly oscillatory function of $b$ and quickly convergent to a certain value. However, this method underestimates the integral value if choosing a small $b$. The red lines and black lines are the integral values computed by using the asymptotic expansion method with $n_{\rm u}=2$ and $7$, respectively, which apparently converge fast with increasing $b$. The purple line in the right panel of Figure~\ref{fig:diff_int_p} shows the integral results computed by using the {\bf Levin's} method, which is basically similar to that of directly integration (green line). To obtain the improper integral value on infinite interval, we may need generalize the method as discussed in Section~\ref{sec:Levin}.

Figures~\ref{fig:point_new}, \ref{fig:point_log} and \ref{fig:NFW_new}, as mentioned in Section~\ref{sec:zpae}, show the comparison of the integral results obtained by using the zero points integral method and the zero points asymptotic expansion method with that obtained by the asymptotic expansion method. Obviously the zero points integral $I(z_k)$ (brown points) and $I_{CS}(z_k)$ (blue points) converge relatively much faster than both the direct integration (green line in Figure~\ref{fig:diff_int_p}) without significant oscillation at the same $b$ and the integral mean $I_C(b)$ (blue line in Figure~\ref{fig:diff_int_p}). However, they converge at significantly larger $b$ compared with that derived by using the {\bf second-order} asymptotic expansion (purple dotted line). The zero points asymptotic expansion method is an excellent choice to compute the diffraction integral. Although it obtains the same integral values with asymptotic expansion at the same $b=z_k$, but it is much more efficient in obtaining the convergent value sine it only needs to calculate a small number of the integral values at zero points. In addition, we can also average these integral values at zero points or use other methods to accelerate the convergence of the integral \cite{NIST:DLMF}.

One has to determine a proper upper limit $b$ by observing the variation of the integral value with increasing $\xi$ (or $x$). If the integral value does not change significantly (within the required precision) with increasing $b$, then it is assumed to be convergent, although the convergence is not completely proved. For the integral mean method, when $b>100$, the diffraction integral is nearly convergent with an error less than $0.1$ for those cases shown in Figures~\ref{fig:point_new} and \ref{fig:NFW_new}. For the asymptotic expansion method with $n_{\rm u}\geq2$, the diffraction integral is basically convergent with an error $<0.001$ when $b>10$ . For $n_{\rm u}=7$, the diffraction integral is convergent with an error $<0.001$ even when $b$ is as small as $2$. The zero points integral $I(z_k)$ is convergent when $k>8$ with an error $<0.001$, and the mean of the zero points integrals $I_{CS}(z_k)$ is convergent when $k>2$ with an error $<0.001$. For the method by combining the zero points integral $I(z_k)$ with the $2$nd-order or $7$th-order asymptotic expansion, their errors are all less than $0.001$. Once adopting $k>3$ ($k>1$), the calculation errors of the combination of $I(z_k)$ with {\bf second(seventh)}-order asymptotic expansion is much less than $10^{-5}$. When the error of $|F|$ is less than $0.001$, the error of GW template is less than $0.001$, which means only if the SNR of signal reach nearly $1000$, two templates can be distinguished \citep{Lindblom:2008cm}.

 Generally, for a larger $w$, it is required to set a larger $wb$ in order to reach the same accuracy for almost all integral methods.
%It costs more time to compute $I(b)$ to the same $b$ if $w$ is higher. 
When $w=100$, or even $10^3$, these methods still work well. The right panel in Figure~\ref{fig:point_log} also shows the relative errors of different integration methods for $w=100$.  However, if $w$ is too large, e.g., $w\gg10^3$, it appears that all integral methods involving Guassian quadrature cannot work effectively simply due to too many oscillations in the integral which may be easily left out by coarse sampling. In theory, {\bf Levin's}  method and Filon method will become more accurate when $w$ is much higher\cite{CHOIch3}, whether it is right to improper integral is to be investigated. 
For very high $w$ like $w>10^3$, if we still want to adopt zero points asymptotic expansion, we may need to use {\bf Levin's} method or Filon-type method to calculate $I(b)$ or $I(z_k)$ on finite interval, not usual Guassian quadrature then use asymptotic expansion to compute this left expansion terms.
Fortunately, when $w>10^3$,
%{\color{red} \bf check whether it is $w\gg1$ or something like $w\gg 10^3$ if it is the value $1$ then why we adopt $w=10$ for figures 2 and 3? }???
the wave optics can be usually well approximated by the geometrical optics\cite{Takahashi_Thesis,2004A&A...423..787T,2018PhRvD..98j4029D}.\footnote{This criterion requires parameter $\xi_0=r_{\rm E}$, where $r_{\rm E}$ is Einstein radius. For NFW model, $\xi_0=r_{\rm s}$ is usually not the Einstein radius but scale radius for convenience. Thus this critical value of $w$ from wave optics to geometrical optics may be much different in such case.} 
The computation of amplification factor in geometrical optics approximation is easy to perform.

The parameter $y$ may not have a significant influence on the of speed of convergence of $F(w,y)$.
However, $|F|$ becomes larger if adopting a smaller $y$, which requires the computation of $F(w,y)$ to reach higher relative errors if it can reach the same absolute error. 
%Because the absolute errors of $F(w,y)$ directly influence the GW waveform.{\color{red} the above sentence is not a good statement. you need to make it more clear???}
Therefore, to reach the same absolute accuracy, it usually requires a larger $b$ for a smaller $y$.

The integral mean method can be used to accelerate the convergence in calculating the diffraction integral, but it is less efficient than the asymptotic expansion method. The asymptotic expansion method leads to a fast convergence in calculating the diffraction integral, especially when choosing $n_{\rm u}=7$ in our cases. However, the asymptotic expansion method still needs to evaluate the integral $I(b)$ which is usually rapidly oscillating. If one only evaluates $I(z_k)$ at zero points, one can obtain the convergent value more efficiently. The calculation errors of the zero points integral method is slightly larger than that of the asymptotic expansion method with $n_{\rm u}=2$ at the same $b=z_k$. However, the zero points integral method enables the removal of rapidly oscillation of the integral value by choosing proper zero points. With this method, the integral values are needed to compute only at a limited number of zero points. This is why it enables efficient estimate of the integral with high accuracy but costing much less computational time. The combination of the zero points integral method and the asymptotic expansion method can have both the advantages of these two methods. The asymptotic expansion method makes use of the information of the integrand derivatives to accelerate the convergence. With this method, $b$ is not necessarily to be set as an extremely large value, as long as the error is within the required precision. The zero points integral method avoids the rapidly oscillation of integral value $I(b)$. Therefore, the zero points asymptotic expansion method is fast and efficient and enables the estimate of the diffraction integral with extremely high accuracy.

The integral mean method, asymptotic expansion method, zero points integral method and so on can also be used to the computation of $2$-dimensional integrals, merely $2$-dimensional integral usually costs much more time than $1$-dimensional integral under the same condition. Take the axial symmetric $1$-dimensional diffraction integral as an example, if we substitute the Bessel function $J_0(z)$ by the integral $I_{\theta_x}(z)$ with respect to angle $\theta_x$, they also can be numerically calculated to obtain the same integral value as the $1$-dimensional diffraction integral. To accelerate the numerical computation of the $2$-dimensional diffraction integral, one may interpolate $I_{\theta_x}(z)$ to obtain an approximate interpolate function $P(z)\approx I_{\theta_x}(z)$. One may have to compute more points of $I_{\theta_x}(z)$ in order to avoid the lost of details of $I_{\theta_x}(z)$. For example, for the point mass model,  $I_{\theta_x}(z)$ is highly oscillatory near $z=0$. If one wants to use interpolation to accelerate the computation of the $2$-dimensional integrals and keep high accuracy at the same time, one needs to sample many more points near $z=0$. With this interpolation, one can compute the secondary integral $I_2(b)\approx\int_0^{b}e^{iwz}P(z)$, also an $1$-dimensional integral, with relatively high efficiency. This $1$-dimension integral can be calculated out by the integral mean method, the asymptotic expansion method, the zero points integral method, or the zero points asymptotic expansion method and so on. We summarize the whole process to quickly compute $2$-dimensional integral as follows:
\begin{enumerate}
\item computing $I_{\theta_x}(z)$ for a number of sampling points, e.g., $10000$ points or so, sampling more points near the oscillatory points to achieving high accuracy;
\item interpolating $I_{\theta_x}(z)$ and obtaining interpolation function $P(z)\approx I_{\theta_x}(z)$;
\item computing $I_2(b)\approx\int_0^{b}e^{iwz}P(z)$ by using one of the methods introduced in this paper, i.e., the integral mean method, the asymptotic expansion method, the zero points integral method, or the zero points asymptotic expansion method, and so on to obtain the estimate value of $I(\infty)$.
\end{enumerate}
Therefore, one may only need to compute two $1$-dimensional integrals, i.e., $I_{\theta_x}(z)$ and $\int_0^{b}e^{iwz}P(z)$, respectively. The consuming time for using such an method to obtain the $2$-dimension diffraction integral at a given accuracy is on the same order of magnitude for that using the $1$-dimensional diffraction integral. 

The {\bf Levin's} method is a common method to compute the integral of highly oscillating function. It transforms the integral problem into the solution of an algebraic equation system. It is feasible for some lens models, such as the SIS lens model and the point mass model. For the NFW lens model, however, it is difficult to calculate by using this method possibly due to the singularity of ${\rm arctanh(1)}$. Although the procedure for the {\bf Levin's} method is a little tedious, it can be used to compute the diffraction integral with some ready-made Mathematic softwares, e.g., the \texttt{NIntegrate} with \texttt{LevinRule} in the \texttt{Mathematica} software directly.

All these integral methods can be used to compute not only the traditional diffraction integral, but also the general diffraction integral (see Equation~\eqref{eq:F_wycos}). 

\section{Conclusions}
\label{sec:concl}

The wave optics may be important for the gravitational lensing of GWs that will be probably detected by future GW detectors since the GW wavelength can be comparable to the Einstein radius of the lens. For the detection of such GW lensing events, it is important to obtain the lensed GW signals accurately and efficiently by calculating the diffraction integrals, of which the integrand is rapidly oscillating. In this paper, we investigate the convergence of the diffraction integrals and find that the traditional diffraction integral, obtained by using the small angle approximation, is usually convergent. Even not convergent under the usual definition, it is also $(C,1)$ summable.
%{\bf We demonstrate a more general diffraction integral expression Equation~\eqref{eq:F_wycos}, its convergency can be well-defined. Although it has only tiny difference with traditional diffraction integral Equation~\eqref{eq:F_wy}, only very-high-SNR detection can distinguish the difference. }
We overview some methods that can be used to calculate the diffraction integral introduced in the literature, such as the asymptotic expansion method, the {\bf Levin's} method, etc. We further introduce several new methods to compute the diffraction integral, such as the integral mean method, the zero points integral method, and a hybrid method by combing the zero points integral method with the asymptotic expansion method, and we compare these new methods with the {\bf Levin's} method and the asymptotic expansion method in terms of the convergence and efficiency. We find that each method has its advantages and disadvantages, and the zero points asymptotic expansion method is probably the most efficient numerical recipe to compute the diffraction integral with the highest accuracy and least computational burden, as it only needs to evaluate the integral values at several zero points of the oscillating integrand. These methods can also be used to $2$-dimensional diffraction integral efficiently and accurately, and the time it takes is roughly on the order of that for computing two $1$-dimensional integrals. 
These efficient numerical integral methods would be important for efficient and fast calculations of a large template bank of lensed GW signals in the wave optics regime, which must be used for the matched filtering search of lensing GW events in the future.

\acknowledgments 
We thank Shun-Sheng Li for helpful discussions.
This work is partly supported by the National Natural Science Foundation of China (Grant No. 11690024, 11873056, 11991052), the Strategic Priority Program of the Chinese Academy of Sciences (Grant No. XDB 23040100), and the National Key Program for Science and Technology Research and Development (Grant No. 2016YFA0400704).

\appendix

\section{The Derivation of the Diffraction Integral in Wave Optics}
\label{sec:deriv}
Assuming GWs propagate under the gravitational potential $U(\boldsymbol{r})$ ($\ll 1$) of a lens object/system, the background spacetime is given by  \citep{1973grav.book.....M}
\begin{equation}
d s^{2}=-(1+2 U) d t^{2}+(1-2 U) d\boldsymbol{r}^{2} \equiv g_{\mu \nu}^{(\mathrm{B})} d x^{\mu} d x^{\nu}.
\label{eq:background}
\end{equation}
We regard the influence of GW on the background spacetime as a linear perturbation $h_{\mu\nu}$, i.e.,
$$g_{\mu \nu}=g_{\mu \nu}^{(\mathrm{B})}+h_{\mu \nu}.$$
Adopting the transverse traceless Lorentz gauge condition $h^{\nu}_{\mu;\nu}=0$, $h^{\mu}_{\mu}=0$, then we have
$$h_{\mu \nu ; \alpha}^{; \alpha}+2 R_{\alpha \mu \beta \nu}^{(\mathrm{B})} h^{\alpha \beta}=0,$$
where the semicolon represents the covariant derivative corresponding to the metric $g^{(\mathrm{B})}_{\mu\nu}$, $R_{\alpha \mu \beta \nu}^{(\mathrm{B})}$ is the Riemann tensor of the background spacetime. If the GW wavelength ($\lambda$) is much less than the   curvature radius of the background spacetime ($\mathcal{R}$), we have 
$$h_{\mu \nu ; \alpha}^{; \alpha}=0.$$
Adopting the eikonal approximation by \cite{1999PhRvD..59h3001B}, GW can be expressed as a form of scalar wave 
$$h_{\mu \nu}=\phi e_{\mu \nu},$$
where $e_{\mu \nu}$ is the GW polarization tensor. Since $U\ll1$, the change of GW polarization tensor is small thus it can be regarded as a constant (see \citep{2019PhRvD.100f4028H}). Therefore, scalar wave is a proper approximation for the cases considered in this paper ($U \ll 1$). The propagation equation of the scalar wave is
\begin{equation}
\partial_{\mu}\left(\sqrt{-g^{(\mathrm{B})}} g^{(\mathrm{B}) \mu \nu} \partial_{\nu} \phi\right)=0,
\label{eq:scalar_pro}
\end{equation}
where $\partial_\mu\equiv \partial/\partial x^\mu$. With the background spacetime Equation~\eqref{eq:background}, Equation~\eqref{eq:scalar_pro} in the frequency domain can be expressed as 
\begin{equation}
\left(\nabla^{2}+\omega^{2}\right) \tilde{\phi}=4 \omega^{2} U \tilde{\phi},
\label{eq:nonhom}
\end{equation}
where $\omega= 2\pi f$ is the circular frequency of GW, $f$ the GW frequency, and $\tilde{\phi}$ the Fourier transform of $\phi$. This equation can be solved by using the Green function method. 

Figure~\ref{fig:illustr} shows the geometrical configuration of the observer-lens-source system. Various parameters involved in such a physical system are described in the figure caption \cite{Takahashi_Thesis}. Adopting the thin lens approximation, $U\approx0$ inside the volume $V$ (see Fig.~\ref{fig:illustr}), then Equation~\eqref{eq:nonhom} can be reduced to the Helmholtz equation
\begin{equation}
\left(\nabla^{2}+\omega^{2}\right) \tilde{\phi}=0.
\label{eq:Helmholtz}
\end{equation}
Green function of a spherical wave $e^{i\omega r}/r$ centering around the observer must satisfy
\begin{equation}
\left(\nabla^{2}+\omega^{2}\right) \frac{e^{i \omega r}}{r}=-4 \pi \delta^{3}(\boldsymbol{r}),
\label{eq:Green}
\end{equation}
where $r$ is the distance from the observer. Combining Equations~\eqref{eq:Helmholtz} and \eqref{eq:Green} together, the scalar field detected by the observer is then
\begin{equation}
\tilde{\phi}_{\rm obs}^{L}=-\frac{1}{4 \pi} \int_{V} d V\left[\tilde{\phi} \nabla^{2} \frac{e^{i \omega r}}{r}-\frac{e^{i \omega r}}{r} \nabla^{2} \tilde{\phi}\right].
\end{equation}
When $R\rightarrow\infty$, $V\rightarrow\infty$, the above volume integral can be transformed into a surface integral on the lens plane $S$ by the Green's theorem
\begin{equation}
\tilde{\phi}_{\rm obs}^{L}=\frac{1}{4 \pi} \iint_{S} d^{2} \xi\left[\tilde{\phi} \frac{\partial}{\partial n} \frac{e^{i \omega r}}{r}-\frac{e^{i \omega r}}{r} \frac{\partial}{\partial n} \tilde{\phi}\right].
\label{eq:Greens}
\end{equation}
The scalar wave is expressed as 
\begin{equation}
\tilde{\phi}=A e^{i S_{\rm P}},
\label{eq:scalar}
\end{equation}
where $A$ and $S_{\rm P}$ are the amplitude and phase, respectively. According to the eikonal approximation, the phase can be written as \citep{Takahashi_Thesis}
\begin{equation}
S_{\rm P}=\omega\left(t_{\rm d}-r\right),
\end{equation}
where $t_{\rm d}$ is the time delay (see its detailed expression in \citep{1992grle.book.....S, 2003ApJ...595.1039T}). After substituting the expression of the scalar wave Equation~\eqref{eq:scalar_pro} into Equation~\eqref{eq:Greens}, we can obtain Equation~\eqref{eq:phi_incl}.

\section{Ces\`{a}ro Summability}
\label{sec:Ces_sum}

According to \cite{NIST:DLMF}, for a series, 
$$s_{n}=\sum_{k=0}^{n} a_{k},$$
we can define Ces\`{a}ro Summability as follow.\\
If
$$
\lim _{n \rightarrow \infty} \frac{s_{0}+s_{1}+\cdots+s_{n}}{n+1}=s,
$$
we declare $s_n$ is $(C,1)$ summable and have
$$
\sum_{n=0}^{\infty} a_{n}=s \quad(C, 1).
$$
For a improper integral over infinite interval such as
$
\int_{-\infty}^{\infty}f(t)dt,
$
if
$$
\lim _{R \rightarrow \infty} \int_{-R}^{R}\left(1-\frac{|t|}{R}\right) f(t) d t=L,
$$
then we declare $\int_{-\infty}^{\infty}f(t)dt$ is $(C,1)$ summable to $L$ or
$$
\int_{-\infty}^{\infty} f(t) d t=L \quad(C, 1).
$$

The oscillatory improper integral in this paper 
$
\int_{0}^{\infty}f(t)dt
$
likes the Fourier integral $\int_0^{\infty}g(t)e^{iwt}dt$ and could be divergent under the usual integral convergence definition. However, it may be $(C,1)$ summable to a finite value $L$, if it satisfies
$$
\lim _{R \rightarrow \infty} \int_{0}^{R}\frac{R-t}{R} f(t) d t=L.
$$
We can prove that this formula is the limit of the integral mean of $\int f(t)dt$ when the upper limit goes to infinity. Under the definition of $(C,1)$ summability,
$$
\int_{0}^{\infty}f(t)dt\equiv\lim_{R\rightarrow\infty}\frac{1}{R}\int_0^{R}dT\int_0^{T}dt f(t) = \lim_{R\rightarrow\infty} \frac{1}{R}\int_{0}^{R}dt\int_{t}^{R}dTf(t)= \lim_{R\rightarrow\infty} \int_{0}^{R}dt\frac{(R-t)}{R}f(t),
$$
where $\frac{\int_0^{R}G(T)dT}{R}$ represents the integral mean of $G(T)$ over $[0,R]$.
In addition, if the integral is convergent to a value $L$ under the usual integral convergence definition, it is also $(C,1)$ summable to the same value $L$ \citep{NIST:DLMF}.

%\nocite{*}
%
%\bibliographystyle{plain}
\bibliography{refer}

\end{document}